\documentclass[12pt,preprint]{aastex}
\usepackage{amsmath,amssymb,graphicx,soul,color}

\usepackage{natbib}

\shorttitle{GRB 110205A}

\slugcomment{Draft version on Feb. 17 2012}

\newcommand{\beq}{\begin{equation}}
\newcommand{\eeq}{\end{equation}}
\newcommand{\bea}{\begin{eqnarray}}
\newcommand{\eea}{\end{eqnarray}}

\def\G{\Gamma}
\def\g{\gamma}

\def\eps{\epsilon}

\begin{document}

\title{Panchromatic observations of the textbook GRB 110205A: constraining physical mechanisms of prompt emission and afterglow$^0$}

\author{ W.~Zheng\altaffilmark{1},
%\email{zwk@umich.edu}
R. F. Shen\altaffilmark{2},
T.    Sakamoto\altaffilmark{3,4,5},
A. P. Beardmore\altaffilmark{6},
M.    De Pasquale\altaffilmark{7},
X. F. Wu\altaffilmark{8,9},
J.    Gorosabel\altaffilmark{10},
Y.    Urata\altaffilmark{11},
S.    Sugita\altaffilmark{12},
B.    Zhang\altaffilmark{8},
A.    Pozanenko\altaffilmark{13},
M.    Nissinen\altaffilmark{14},
D. K. Sahu\altaffilmark{15},
M.    Im\altaffilmark{16},
T. N. Ukwatta\altaffilmark{17},
M.    Andreev\altaffilmark{18,19},
E.    Klunko\altaffilmark{20},
A.    Volnova\altaffilmark{21},
C. W. Akerlof\altaffilmark{1},
P.    Anto\altaffilmark{15},
S. D. Barthelmy\altaffilmark{5},
A.    Breeveld\altaffilmark{7},
U.    Carsenty\altaffilmark{22},
S.    Castillo-Carri\'on\altaffilmark{23},
A. J. Castro-Tirado\altaffilmark{10},
M. M. Chester\altaffilmark{24},
C. J. Chuang\altaffilmark{11},
R.    Cunniffe\altaffilmark{10},
A.    De Ugarte Postigo\altaffilmark{25},
R.    Duffard\altaffilmark{10},
H.    Flewelling\altaffilmark{26},
N.    Gehrels\altaffilmark{5},
T.    G\"uver\altaffilmark{27},
S.    Guziy\altaffilmark{10},
V. P. Hentunen\altaffilmark{14},
K. Y. Huang\altaffilmark{28},
M.    Jel\'{\i}nek\altaffilmark{10},
T. S. Koch\altaffilmark{24},
P.    Kub\'anek\altaffilmark{10},
P.    Kuin\altaffilmark{7},
T. A. McKay\altaffilmark{1},
S.    Mottola\altaffilmark{22},
S. R. Oates\altaffilmark{7},
P.    O'Brien\altaffilmark{6},
M.    Ohno\altaffilmark{29}
M. J. Page\altaffilmark{7},
S. B. Pandey\altaffilmark{30},
C.    P\'erez del Pulgar\altaffilmark{23},
W.    Rujopakarn\altaffilmark{31},
E.    Rykoff\altaffilmark{32},
T.    Salmi\altaffilmark{14},
R.    S\'anchez-Ram\'{\i}rez\altaffilmark{10},
B. E. Schaefer\altaffilmark{33},
A.    Sergeev\altaffilmark{18,19},
E.    Sonbas\altaffilmark{5,34,35},
A.    Sota\altaffilmark{10},
J. C. Tello\altaffilmark{10},
K.    Yamaoka\altaffilmark{36},
S. A. Yost\altaffilmark{37},
F.    Yuan\altaffilmark{38,39}
}

\altaffiltext{0} {Correspondence authors : W. Zheng, zwk@umich.edu; R. Shen, rfshen@astro.utoronto.ca ; B. Zhang, zhang@physics.unlv.edu}
\altaffiltext{1} {University of Michigan, 450 Church Street, Ann Arbor, MI, 48109, USA}
\altaffiltext{2}{Department of Astronomy and Astrophysics, University of Toronto, Toronto, Ontario, M5S 3H4, Canada}
\altaffiltext{3}{Center for Research and Exploration in Space Science and Technology (CRESST), NASA Goddard Space Flight Center, Greenbelt, MD 20771}
\altaffiltext{4}{Joint Center for Astrophysics, University of Maryland, Baltimore County, 1000 Hilltop Circle, Baltimore, MD 21250}
\altaffiltext{5}{NASA Goddard Space Flight Center, Greenbelt, MD 20771}
\altaffiltext{6}{Department of Physics and Astronomy, University of Leicester, Leicester, LE1 7RH, UK}
\altaffiltext{7}{Mullard Space Science Laboratory, University College London , Holmbury Road, Holmbury St. Mary, Dorking RH5 6NT, UK}
\altaffiltext{8}{Department of Physics and Astronomy, University of Nevada Las Vegas, Las Vegas, NV 89154, USA}
\altaffiltext{9}{Purple Mountain Observatory, Chinese Academy of Sciences, Nanjing 210008, China}
\altaffiltext{10}{Instituto de Astrof\'{\i}sica de Andaluc\'{\i}a (IAA-CSIC), 18008 Granada, Spain}
\altaffiltext{11}{Institute of Astronomy, National Central University, Chung-Li 32054, Taiwan}
\altaffiltext{12}{EcoTopia Science Institute, Nagoya University, Furo-cho, chikusa, Nagoya 464-8603, Japan}
\altaffiltext{13}{Space Research Institute (IKI), 84/32 Profsoyuznaya st., Moscow 117997, Russia}
\altaffiltext{14}{Taurus Hill Observatory, H\"ark\"am\"aentie 88, 79480 Kangaslampi, Finland}
\altaffiltext{15}{CREST, Indian Institute of Astrophysics, Koramangala, Bangalore, 560 034, India}
\altaffiltext{16}{Center for the Exploration of the Origin of the Universe, Department of Physics \& Astronomy, FPRD, Seoul National University, Shillim-dong, San 56-1, Kwanak-gu, Seoul, Korea}
\altaffiltext{17}{Department of Physics and Astronomy, Michigan State University, East Lansing, MI 48824, USA}
\altaffiltext{18}{Terskol Branch of Institute of Astronomy of RAS, Kabardino-Balkaria Republic 361605, Russian Federation}
\altaffiltext{19}{International Centre of Astronomical and Medico-Ecological Research of NASU, 27 Akademika Zabolotnoho st,. 03680 Kyiv, Ukraine}
\altaffiltext{20}{Institute of Solar-Terrestrial Physics, Lermontov st., 126a, Irkutsk 664033, Russia}
\altaffiltext{21}{Sternberg Astronomical Institute, Moscow State University, Universitetsky pr., 13, Moscow 119992, Russia}
\altaffiltext{22} {Institute of Planetary Research, DLR, 12489 Berlin, Germany}
\altaffiltext{23} {Universidad de M\'alaga, Campus de Teatinos, M\'alaga, Spain.}
\altaffiltext{24} {Department of Astronomy \& Astrophysics, Penn State University, 525 Davey Laboratory, University Park, PA 16802, USA}
\altaffiltext{25} {Dark Cosmology Centre, Niels Bohr Institute, University of Copenhagen, Juliane Maries Vej 30, DK-2100 Copenhagen O, Denmark}
\altaffiltext{26} {Institute for Astronomy, 2680 Woodlawn Ave, Honolulu, HI, 96822}
\altaffiltext{27} {Department of Astronomy, University of Arizona, 933 N. Cherry Ave., Tucson, AZ 85721}
\altaffiltext{28} {Academia Sinica Institute of Astronomy and Astrophysics, Taipei 106, Taiwan}
\altaffiltext{29} {Japan Aerospace Exploration Agency, Institute of Space and Astronautical Science, 3-1-1 Yoshinodai, Chuo-ku, Sagamihara, Kanagawa 252-5210, Japan}
\altaffiltext{30} {Aryabhatta Research Institute of observational sciencES (ARIES), Manora Peak, Naini Tal, 263129, India}
\altaffiltext{31} {Steward Observatory, The University of Arizona, Tucson, AZ 85721, USA}
\altaffiltext{32} {E. O. Lawrence Berkeley National Lab, 1 Cyclotron Rd., Berkeley CA, 94720, USA}
\altaffiltext{33} {Department of Physics and Astronomy, Louisiana State University, Baton Rouge, LA, 70803, USA}
\altaffiltext{34} {University of Ad{\i}yaman, Department of Physics, 02040 Ad{\i}yaman, Turkey}
\altaffiltext{35} {Universities Space Research Association, 10211 Wincopin Circle, Suite 500, Columbia, MD 21044-3432, USA}
\altaffiltext{36} {Department of Physics and Mathematics, Aoyama Gakuin University, 5-10-1 Fuchinobe, Chuo-ku, Sagamihara, Kanagawa 252-5258, Japan}
\altaffiltext{37} {Department of Physics, College of St. Benedict, Collegeville, MN 56321, USA}
\altaffiltext{38} {Research School of Astronomy and Astrophysics, The Australian National University, Weston Creek, ACT 2611, Australia}
\altaffiltext{39} {ARC Centre of Excellence for All-sky Astrophysics (CAASTRO)}

\shortauthors{Zheng W. et al. 2010}

\begin{abstract}
We present a comprehensive analysis of a bright, long duration
(T$_{90}$ $\sim$ 257 s) GRB 110205A
at redshift $z= 2.22$. The optical prompt emission was detected by
$Swift$/UVOT,
ROTSE-IIIb and BOOTES telescopes when the GRB was still radiating in
the $\gamma$-ray band, with optical lightcurve showing correlation
with $\gamma$-ray data. Nearly 200 s
of observations were obtained simultaneously from optical, X-ray to
$\gamma$-ray (1 eV - 5 MeV), which makes it one of the exceptional
cases to study the broadband spectral energy distribution
during the prompt emission phase. In particular, 
we clearly identify, for the first
time, an interesting two-break energy spectrum, roughly
consistent with the standard synchrotron emission model in the
fast cooling regime.
Shortly after prompt emission ($\sim$ 1100 s), a
bright ($R$ = 14.0) optical emission hump with very steep rise
($\alpha$ $\sim$ 5.5) was observed which we interpret as the
the reverse shock emission. It is the first time that the
rising phase of a reverse shock component has been closely observed.
The full optical and X-ray afterglow lightcurves
can be interpreted within the standard reverse shock (RS) +
forward shock (FS) model.
In general, the high quality prompt and afterglow data
allow us to apply the standard fireball model to extract
valuable information including the
radiation mechanism (synchrotron), radius of prompt emission
($R_{\rm GRB}\sim 3\times 10^{13}$ cm), initial Lorentz factor of
the outflow ($\Gamma_0 \sim 250$), the composition of
the ejecta (mildly magnetized), as well as the collimation
angle and the total energy budget.
\end{abstract}


\section{Introduction}
Gamma-Ray Bursts (GRBs) are extremely luminous explosions in the
universe. A standard fireball model (e.g. Rees \& M\'esz\'aros 1992,
1994; M\'esz\'aros \& Rees 1997; Wijers, Rees \& M\'esz\'aros 1997;
Sari, Piran \& Narayan 1998; see e.g. Zhang \& M\'esz\'aros 2004,
M\'esz\'aros 2006 for reviews)
has been developed following their discovery in 1973
(Klebesadel, Strong \& Olson 1973) to explain their observational
nature. Generally, the prompt emission can be modeled as 
originating from internal shocks or the photosphere of the fireball
ejecta or magnetic dissipation from a magnetically dominated jet, 
while the afterglow emission originates
from external shocks that may include both forward shock and
reverse shock components (M\'esz\'aros \& Rees 1997, 1999; Sari \&
Piran 1999).

The leading radiation mechanisms of the GRB prompt emission are synchrotron
radiation, synchrotron self-Compton (SSC), and Compton upscattering of a
thermal seed photon source (e.g. Zhang 2011 for a review). All these
mechanisms give a ``non-thermal" nature to the GRB prompt spectrum.
Observationally, the prompt spectrum in the $\gamma$-ray band
can be fit with a smoothly broken power-law called the Band function
(Band et al. 1993).
Since this function is characterized by a single break energy,
it can not adequately fit the spectrum if
the spectral distribution is too complex. For example, the synchrotron mechanism
predicts an overall power-law spectrum characterized by
several break frequencies: $\nu_a$ (self-absorption
frequency), $\nu_m$ (the frequency of minimum electron injection energy), and $\nu_c$ (cooling
frequency) (Sari \& Esin 2001). However, due to instrumental and
observational constraints, it is almost impossible to cover the
entire energy range and re-construct the
prompt spectrum with all three predicted break points. Thus, despite its
limited number of degrees of freedom, the Band function is an empirically good description 
for most GRBs.

The $Swift$ mission (Gehrels et al. 2004), thanks to its rapid and
precise localization capability, performs simultaneous
observations in the optical to $\gamma$-ray
bands, allowing broadband observations of the prompt phase much more frequently than previous GRB probes. This energy range may also
span up to 6 orders of magnitude (e.g. GRB 090510, Abod et al. 2009; De Pasquale et al. 2010) 
if a GRB is observed by both the $Swift$  and $Fermi$ satellites (Atwood et al. 2009; Meegan
et al. 2009). Some prompt observations have shown signatures of a
synchrotron spectrum from the break energies ($\nu_m$,$\nu_c$) (e.g. GRB
080928, Rossi et al. 2011). Prompt optical observations can also be
used to constrain the self-absorption frequency $\nu_a$ (Shen \&
Zhang 2009), but, so far, no GRB has been observed clearly with more
than two break energies in the prompt spectrum. Meanwhile, early
time observations in the optical band provide a greater chance to detect
reverse shock emission which has only been observed for a few bursts
since the first detection in GRB 990123 (Akerlof et al. 1999).

Here we report on the analysis of the long duration GRB 110205A
triggered by the $Swift$/BAT (Barthelmy et al. 2005). Both prompt and
afterglow emissions are detected with good data sampling. Broadband
energy coverage over 6 orders of magnitude (1 eV - 5 MeV) during prompt
emission makes this GRB a rare case from which we can study the prompt spectrum
in great detail. Its bright optical ($R$ = 14.0 mag) and X-ray
afterglows allow us to test the different external shock models and to
constrain the physical parameters of the fireball model.

Throughout this paper we adopt a standard cosmology model with
H$_0$ = 71 km s$^{-1}$ Mpc$^{-1}$, $\Omega_M$ = 0.27 and
$\Omega_{\Lambda}$ = 0.73. We use the usual power-law representation
of flux density F($\nu$) $\propto$ t$^{\alpha}\nu^{-\beta}$ for the
further analysis. All errors are given at the 1 $\sigma$ confidence
level unless otherwise stated.

\section{Observations and Data Reductions}
\subsection{Observations}
At 02:02:41 UT on Feb. 5, 2011 (T$_0$), the $Swift$/BAT triggered and
located GRB 110205A (trigger=444643, Beardmore et al. 2011). The
BAT light curve shows many overlapping peaks with a
general slow rise starting at T$_0$-120 s, with the highest peak at
T$_0$+210 s, and ending at T$_0$+1500 s. T$_{90}$ (15-350 keV) is
257 $\pm$ 25 s (estimated error including systematics). GRB 110205A
was also detected by WAM (Sugita et al. 2011, also included in
our analysis) onboard $Suzaku$ (Yamaoka et al. 2009) and Konus-Wind (Golenetskii et al. 2011; Pal'shin 2011) in the $\gamma$-ray
band. A bright, uncatalogued X-ray afterglow was promptly identified
by XRT (Burrows et al. 2005a) 155.4 s after the burst (Beardmore et
al. 2011). The UVOT (Roming et al. 2005) revealed an optical afterglow
164 s after the burst at location RA(J2000) = 10$^h$58$^m$31$^s$.12,
DEC(J2000) = +67$^{\circ}$31'31".2 with a 90\%-confidence error
radius of about 0.63 arc second (Beardmore et al. 2011), which was
later seen to re-brighten (Chester and Beardmore 2011).

ROTSE-IIIb, located at the McDonald Observatory, Texas, responded to GRB
110205A promptly and confirmed the optical afterglow (Schaefer et
al. 2011). The first image started at 02:04:03.4 UT, 82.0 s after
the burst (8.4 s after the GCN notice time). The optical afterglow
was observed to re-brighten dramatically to ~14.0 mag $\sim$
1100 s after the burst, as was also reported by other groups
(e.g. Klotz et al. 2011a,b; Andreev et al. 2011). ROTSE-IIIb continued
monitoring the afterglow until it was no longer detectable,
~1.5 hours after the trigger.

Ground-based optical follow-up observations were also performed by
different groups with various instruments, some of them are presented by
Cucchiara et al. (2011) and Gendre et al. (2011). In this paper, the
optical data includes: Global Rent-a-Scope (GRAS) 005 telescope at New Mexico (Hentunen et al. 2011),
1-m telescope at Mt. Lemmon Optical Astronomy Observatory (LOAO; Im \& Urata 2011, Lee et al. 2010),
Lulin One-meter Telescope (LOT; Urata et al. 2011), 0.61-m Lightbuckets rental telescope LB-0001 in Rodeo,
NM (Ukwatta et al. 2011), 2-m Himalayan Chandra Telescope (HCT; Sahu et al. 2011),
Zeiss-600 telescope at Mt. Terskol observatory (Andreev et al. 2011),
1.6-m AZT-33IK telescope at Sayan Solar observatory, Mondy (Volnova et al. 2011) as well as Burst Observer and Optical Transient Exploring
System (BOOTES 1 and 2 telescopes), 1.23-m and 2.2-m telescope at Calar Alto Observatory and 1.5m OSN telescope
which are not reported in the GCNs.

The redshift measurement of GRB 110205A was reported by three
independent groups with $z$ = 1.98 (Silva et al. 2011), $z$ = 2.22
(Cenko et al. 2011; Cucchiara et al. 2011) and $z$ = 2.22 (Vreeswijk et al. 2011). Here we
adopt $z$ = 2.22 for which two observations are in very close agreement.

\subsection{Data reductions}
The data from $Swift$ and $Suzaku$, including UVOT, XRT, BAT and WAM
(50 keV - 5 MeV), were processed with the standard HEAsoft software
(version 6.10). The BAT and XRT data were further automatically
processed by the Burst Analyser
pipeline\footnote{http://www.swift.ac.uk/burst$\_$analyser/}
(Evans et al. 2007, 2009, 2010),
with the light curves background subtracted. For the
XRT data, Windowed Timing (WT) data and Photon Counting (PC) data were
processed separately. Pile-up corrections were applied if necessary,
especially at early times when the source was very bright. The UVOT
data were also processed with the standard procedures. A count rate was
extracted within a radius of 3 or 5 arcseconds depending on the
source brightness around the best UVOT
coordinates. The data in each filter were binned with $\delta$t/t = 0.2
and then converted to flux density using a UVOT
GRB spectral model (Poole et al. 2008; Breeveld et al 2010, 2011).

For the ground-based optical data, different methods were used for
each instrument. For the ROTSE data, the raw images were processed
using the standard ROTSE software pipeline. Image co-adding was
performed if necessary to obtain a reasonable signal-to-noise ratio.
Photometry was then extracted using the method described in Quimby et
al. (2006). Other optical data were processed using the standard
procedures provided by the IRAF\footnote{IRAF is distributed by NOAO,
which is operated by AURA, Inc., under cooperative agreement with
NSF.} software. A differential aperture photometry was applied with the
DAOPHOT package in IRAF. Reference stars were calibrated using the
photometry data from SDSS (Smith et al. 2002). $Clear (C)$ band
data were calibrated to $R$ band.

The spectral fitting, including WAM, BAT, XRT and optical data, was
performed using Xspec (version 12.5). We constructed a set of prompt
emission spectra over 9 time intervals
during which the optical data were available. Data from each
instrument were re-binned to the same time intervals. In the afterglow phase, SEDs in 4
different epochs were constructed when we have the best coverage of
multi-band data from optical to X-rays: 550 s, 1.1 ks, 5.9 ks
and 35 ks. All the spectral fittings were carried out under Xspec
using $\chi^2$ statistics, except the 1.1 ks SED of the afterglow,
for which C-statistics was used.

\section{Multi-Wavelength Data Analysis}
\subsection{Broad-band prompt emission, from optical to
$\gamma$-rays\label{sec:analysis_prompt}}
Thanks to its long duration (T$_{90}$ = 257 $\pm$ 25 s), GRB
110205A was also detected by XRT and UVOT during the prompt emission
phase starting from 155.4 s and 164 s after the trigger, respectively.
Both XRT and UVOT obtained nearly 200 s of high quality and
well sampled data during the prompt phase. ROTSE-IIIb  and BOOTES
also detected the optical prompt emission 82.0 s and 102 s after the
trigger, respectively. Together with the $\gamma$-ray data collected by
BAT (15 keV - 150 keV) and $Suzaku$/WAM (50 keV - 5 MeV),
these multi-band prompt emission data cover 6 orders of magnitude
in energy, which allow us to study the temporal and spectral
properties of prompt emission in great detail.
%\begin{figure}[!]
\begin{figure}[!hbp]
\centering
   \includegraphics[width=.95\textwidth]{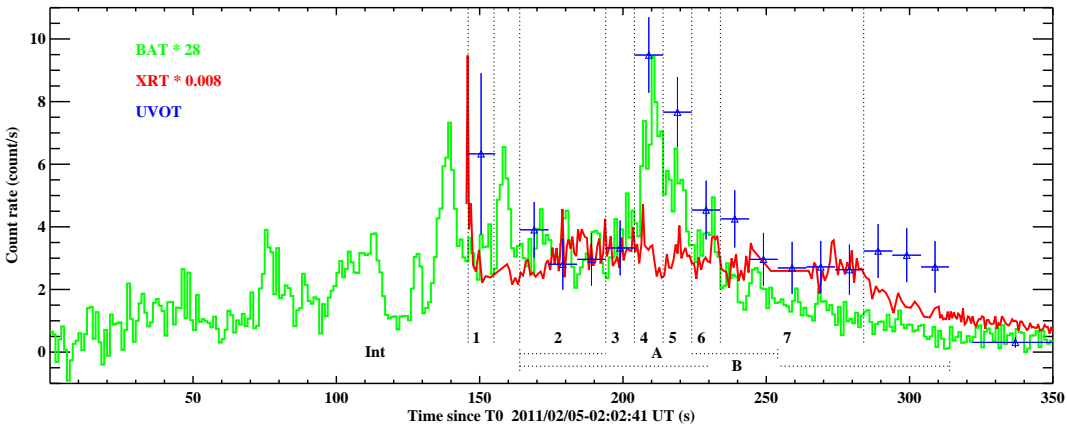}
   \caption{Prompt light curves of GRB 110205A from $Swift$ BAT ($green$),
   XRT($red$) and UVOT($blue$). Arbitrary scale. For the UVOT data,
   the first data point is in $v$, but has been normalized to $white$ band,
   the rest of the data are in $white$. The vertical lines partition 9 intervals
   for constructing the prompt spectra, where Int A and B are averaged ones. \label{fig:110205A_prompt_lc}}
\end{figure}

Figure \ref{fig:110205A_prompt_lc} shows the prompt light curves from
$Swift$ BAT ($green$), XRT ($red$) and UVOT ($blue$). The BAT light curve
shows multiple peaks until at least T$_0$+300 s with a peak count rate
at T$_0$+210 s. The XRT
data show a decay phase from a very bright peak at
the start of XRT observations, followed by smaller peaks with
complicated variability. The UVOT observations were performed mainly in the $white$
band, except for the first point that was observed in the $v$ band
but has been normalized to $white$ using the late time UVOT data.
The UVOT light curve shows only two
major peaks. The first small peak (146 - 180 s) shows weak correlation with the BAT.
After $\sim$40 s, it re-brightens to its
second and brightest peak at 209 s, coinciding with
the brightest $\gamma$-ray peak in the BAT light curve. Overall, the
optical data are smoother, and trace the BAT data better than
the XRT data.

Several vertical lines shown in Figure 1 partition the light curve into 9 different 
intervals according to the UVOT significance criterion to obtain time-resolved 
joint-instrument spectral analysis using the XRT, the BAT and the WAM data. Since
the prompt emission of GRB 110205A is observed by multiple 
instruments, the systematic uncertainty among the 
instruments to perform the joint analysis must be carefully understood.

The energy response function of 
XRT has been examined by the observations of supernova remnants and AGNs with 
various X-ray missions such as {\it Suzaku} and {\it XMM-Newton}.  According to 
the simultaneous observation of Cyg X-1 between XRT (WT mode) and 
{\it Suzaku}/XIS,\footnote{http://swift.gsfc.nasa.gov/docs/heasarc/caldb/swift/docs/xrt/SWIFT-XRT-CALDB-09\_v16.pdf} the photon index and the observed flux agree within $\sim$5\% 
and $\sim$15\% respectively.  The spectral calibration of BAT has been based on Crab nebula 
observations at various boresight angles. The photon index and the flux are 
within $\sim$5\% and $\sim$10\% of the assumed Crab values based on Rothschild et al. (1998)
and Jung et al. (1989).  Similarly, the WAM energy response has been investigated using the Crab 
spectrum collected by the Earth occultation technique (Sakamoto et al. 2011a).  The spectral 
shape and its normalization are consistent within 10-15\% with the result of the {\it INTEGRAL} 
SPI instrument (Sizun et al. 2004). The cross-instrument calibration between BAT and WAM has been 
investigated deeply by Sakamoto et al. (2011b) using simultaneously observed bright GRBs.  
According to this work,
%the low-energy photon index of the Band function in both the BAT and WAM is systematically 
%steeper by 0.2-0.3 comparing to that derived from the Konus-Wind data.
%The BAT-WAM joint fit $E_{\rm peak}$ is systematically higher by 10-15\% comparing to that of the Konus-Wind.  
the normalization of the BAT-WAM joint fit agrees within 10-15\% to the BAT data.  
The cross-instrument calibration between XRT (WT mode) and BAT has been investigated by the 
simultaneous observation of Cyg X-1.  Both the spectral shape and the flux agree within 
5-10\% range between XRT and BAT.\footnote{The presentation in the 2009 {\it Swift} conference: 
http://www.swift.psu.edu/swift-2009/}  In summary, based on the single instrument and the 
cross-instrument calibration effort, the systematic uncertainty among XRT, BAT and WAM should be 
within 15\% in both the spectral shape and its normalization of the spectrum.
To accomodate systematic uncertainties, we include a multiplication factor in the
range 0.85 to 1.15 for the flux normalization for each instrument.

We first applied the spectral analysis to the time-average interval B (intB, see
Figure \ref{fig:110205A_prompt_lc} for interval definition) and
find that the photon index in a simple power-law model derived by the individual instrument 
differs significantly. The photon indices derived by the XRT, the BAT and the WAM spectra 
are $-1.12 \pm 0.03$, $-1.71 \pm 0.04$ and $-2.27_{-0.27}^{+0.22}$, respectively (also listed in
Table \ref{tab:prompt_spec_par}, spectral fitting errors in Table \ref{tab:prompt_spec_par}
and in this section are given in 90\% confidence).   
Since these differences are significantly larger than the systematic uncertainty 
associated with the instrumental 
cross-calibration (the systematic error in the photon index is $\pm$0.3 for the worst 
case as discussed above), the apparent change of spectral slope is very likely 
intrinsic to the GRB. Thus, the observed broad-band spectrum requires two breaks 
to connect the XRT, the BAT and the WAM data.

According to the GRB synchrotron emission model, the overall
spectrum should be a broken power-law
characterized by several break frequencies (e.g. the self-absorption
frequency $\nu_a$, the cooling frequency $\nu_c$, and the frequency
of minimum electron injection energy $\nu_m$).
However, the well
known Band function, which only includes one break energy, cannot
represent the shape of the more complex spectrum of this particular event, therefore we
extended the analysis code, Xspec, to include two additional spectral functions.
The first one is a double ``Band'' spectrum with two spectral
breaks, which was labelled $bkn2band$:
\begin{equation}    \label{eq:bkn2band_F}
f(E) = \begin{cases}
AE^{\alpha}exp(-E/E_0), ~~~~~ if ~E\le\frac{E_0E_1}{E_1-E_0}(\alpha-\beta)\\
  \\
A[\frac{E_0E_1}{E_1-E_0}(\alpha-\beta)]^{\alpha-\beta}exp(\beta-\alpha)E^{\beta}exp(-E/E_1),  \\
if ~\frac{E_0E_1}{E_1-E_0}(\alpha-\beta) < E \le (\beta-\gamma)E_1  \\
  \\
A[\frac{E_0E_1}{E_1-E_0}(\alpha-\beta)]^{\alpha-\beta}exp(\beta-\alpha)\cdot \\
[(\beta-\gamma)E_1]^{\beta-\gamma}exp(\gamma-\beta)E^{\gamma}, \\
if ~E>(\beta-\gamma)E_1&
\end{cases}
\end{equation}
where A is the normalization at 1 keV in unit of ph cm$^{-2}$ s$^{-1}$ keV$^{-1}$,
$\alpha$, $\beta$, and $\gamma$ are the photon indices of the three power law
segments, and $E_0$ and $E_1$ are the two break energies.
However, when fitting the spectrum using this
new $bkn2band$ model,
the third power law index $\gamma$, is poorly constrained
mainly due to the poor statistics in the high energy WAM data above 400 keV. For this reason,
the second new model, $bandcut$, replaces the third power-law
component with an exponential cutoff \footnote{Both
$bkn2band$ and $bandcut$ new models can be downloaded from the following web
page: http://asd.gsfc.nasa.gov/Takanori.Sakamoto/personal/. They can be used to
fit future GRBs with similar characteristics.}:
\begin{equation}    \label{eq:bandcut_F}
f(E) = \begin{cases}
AE^{\alpha}exp(-E/E_0), ~~~~~ if ~E\le\frac{E_0E_1}{E_1-E_0}(\alpha-\beta)\\
  \\
A[\frac{E_0E_1}{E_1-E_0}(\alpha-\beta)]^{\alpha-\beta}exp(\beta-\alpha)E^{\beta}exp(-E/E_1),  \\
if ~E>\frac{E_0E_1}{E_1-E_0}(\alpha-\beta)
\end{cases}
\end{equation}

%\begin{figure}[!]
\begin{figure}[!hbp]
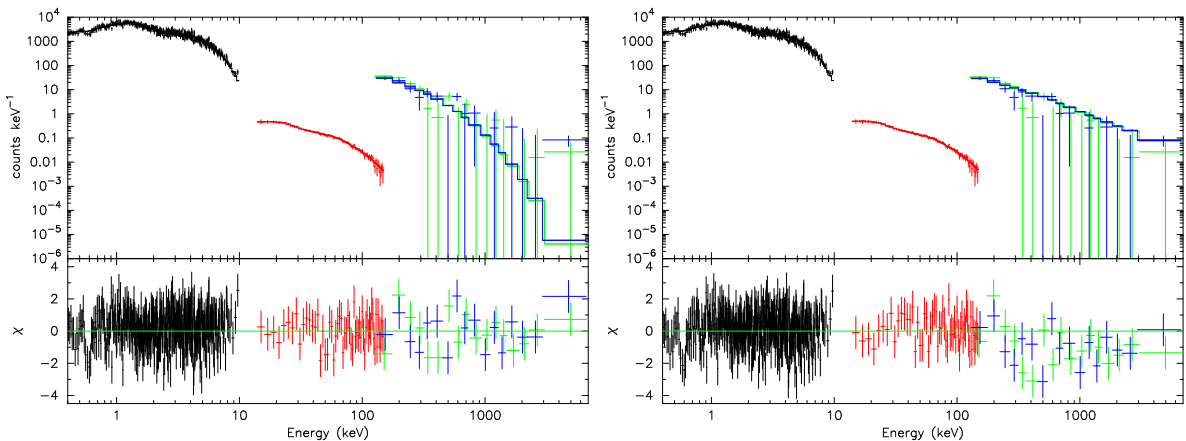

\centering
    \includegraphics[width=.35\textwidth,angle=-90]{f2a.ps}
    \includegraphics[width=.35\textwidth,angle=-90]{f2b.ps}
\caption{The XRT (black), BAT (red) and WAM (green and blue) count spectra of intB  
with the {\it bandcut} model (left panel) and the standard Band function (right 
panel) fit. The improvement using $bandcut$ model can be easily seen in the residual panel at the
bottom. \label{fig:intA_spectrum}}
\end{figure}

Note that the exponential cutoff in the new $bandcut$ model introduces a second break, E1,
although different from the break in a doubly broken power-law model 
or $bkn2band$ model. The new $bandcut$ model can well characterize
the prompt emission spectra of GRB 110205A. Hereafter,
we use ``two-break energy spectrum" to represent the $bandcut$ model spectrum.
Figure \ref{fig:intA_spectrum} shows the XRT, BAT and WAM joint fit spectral analysis 
of intB based on the {\it bandcut} model and the Band function fit.  The systematic residuals 
from the best fit Band function are evident especially in the WAM data.  The best fit parameters 
based on the {\it bandcut} model are $\alpha = -0.50_{-0.08}^{+0.09}$, $E_{0} = 5.0_{-0.8}^{+1.1}$ keV, 
$\beta = -1.54_{-0.09}^{+0.10}$ and $E_{1} = 333_{-118}^{+265}$ keV ($\chi^{2}$/dof = 529.8/503).  
On the other hand, the best fit parameters based on the Band function are $\alpha = -0.59_{-0.08}^{+0.06}$, 
$\beta = -1.72_{-0.03}^{+0.01}$  and $E_{\rm peak} = 9.5_{-0.9}^{+1.5}$ keV ($\chi^{2}$/dof = 575.7/504).  
Therefore, $\Delta\chi^{2}$ ($\equiv \chi^{2}_{Band} - \chi^{2}_{bandcut}$) between the Band function 
and the {\it bandcut} model is 45.9 for 1 dof.  
To quantify the significance of this improvement, we performed 10,000 spectral simulations assuming 
the best fit Band function parameters by folding the energy response functions and the background 
data of XRT, BAT and WAM.  Then, we determine how many cases the {\it bandcut} model fit gives $\chi^{2}$ 
improvements of equal 
or greater than $\Delta\chi^{2}$ = 45.9 for 1 dof over the Band function fit.  We found equal or 
higher improvements in none of the simulated spectra out of 10,000.  Thus, the chance probability of having 
an equal or higher $\Delta\chi^{2}$ of 45.9 with the {\it bandcut} when the parent distribution is 
actually the Band function is $<$0.01\%. A caveat for this simulation is that 
the statistical improvement of the joint fit may be not as high as this simulation indicates if 
the calibration uncertainties among the instruments are included.

The same method is then applied to perform the joint spectral fitting to the remaining time
intervals. Table \ref{tab:prompt_spec_par} shows the best fit results from
the $bandcut$ model.
This is the first time when two spectral breaks are clearly identified in
the prompt GRB spectra. The two breaks are consistent with the expectation
of the broken power law synchrotron spectrum (see discussion in
\S\ref{sec:modeling_prompt}).

%\begin{figure}[!]
\begin{figure}[!hbp]
\centering
   \includegraphics[width=.49\textwidth, angle=0]{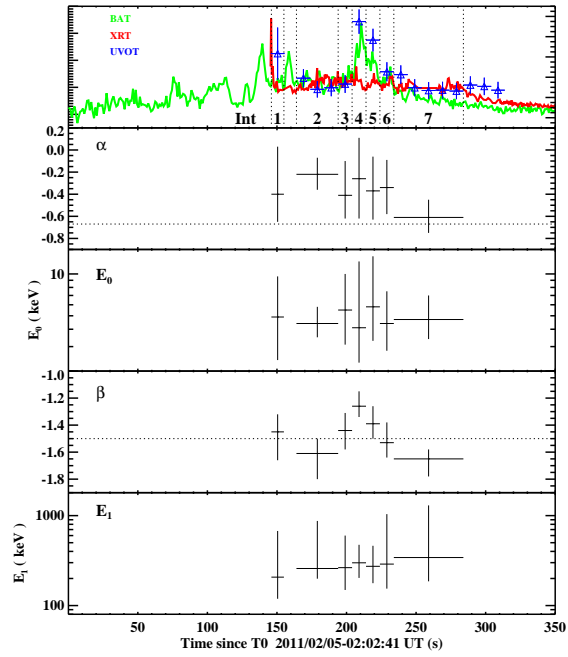}
   \caption{Time evolution of different parameters from the time-resolved
   spectral fitting of GRB 110205A. Top panel has the same key as 
   Figure \ref{fig:110205A_prompt_lc}. The two horizontal dotted lines in $\alpha$
   and $\beta$ panels are the predicted synchrotron photon index value for fast
   cooling phase from the standard GRB fireball model. Errors are given at 90\%
   confidence except the top panel. \label{fig:prompt_spec_par}}
\end{figure}

\clearpage
\begin{deluxetable}{ccccccccccc}
 \tabcolsep 0.4mm
 \tablewidth{0pt}
 \tablecaption{Best spectral fit result from XRT, BAT and WAM joint fitting with $bandcut$ model. The last three columns show the photon index from each single instrument fitting. Errors are given in 90\% confidence.\label{tab:prompt_spec_par}}
  \tablehead{\colhead{Int} & \colhead{t$_1$} & \colhead{t$_2$} & \colhead{$\alpha$} & \colhead{E$_0$} & \colhead{$\beta$}  & \colhead{E$_1$}  & \colhead{$\chi^2$/dof }  & \colhead{XRT} & \colhead{BAT} & \colhead{WAM} }
\startdata
int1 & 146 & 155 & -0.40$_{-0.25}^{+0.43}$ & 4.9$_{-2.5}^{+4.7}$ & -1.45$_{-0.21}^{+0.13}$ & 207$_{-88}^{+468}$   & 126.4/137=0.92 &   -1.05$_{-0.11}^{+0.11}$ & -1.68$_{-0.09}^{+0.09}$ & -2.24$_{-0.70}^{+0.48}$ \\ %1.07            int1
int2 & 164 & 194 & -0.22$_{-0.14}^{+0.15}$ & 4.4$_{-0.9}^{+1.4}$ & -1.61$_{-0.19}^{+0.11}$ & 258$_{-59}^{+615}$ & 269.7/256=1.05 &     -0.94$_{-0.06}^{+0.05}$ & -1.81$_{-0.06}^{+0.06}$ & -2.72$_{-0.84}^{+0.63}$ \\ %1.00            int2-4
int3 & 194 & 204 & -0.41$_{-0.21}^{+0.31}$ & 5.5$_{-2.4}^{+4.5}$ & -1.44$_{-0.14}^{+0.13}$ & 264$_{-115}^{+336}$   & 138.5/145=0.96 &  -0.99$_{-0.11}^{+0.10}$ & -1.62$_{-0.08}^{+0.08}$ & -2.14$_{-0.98}^{+0.57}$ \\ %1.11            int5
int4 & 204 & 214 & -0.26$_{-0.36}^{+0.37}$ & 4.1$_{-1.8}^{+8.2}$ & -1.26$_{-0.08}^{+0.11}$ & 299$_{-96}^{+175}$ & 107.6/146=0.74 &     -0.91$_{-0.10}^{+0.09}$ & -1.45$_{-0.06}^{+0.05}$ & -2.22$_{-0.25}^{+0.20}$ \\ %0.93            int6
int5 & 214 & 224 & -0.37$_{-0.26}^{+0.31}$ & 5.8$_{-2.5}^{+7.6}$ & -1.39$_{-0.11}^{+0.13}$ & 273$_{-96}^{+189}$ & 144.5/144=1.00 &     -0.87$_{-0.10}^{+0.09}$ & -1.58$_{-0.06}^{+0.06}$ & -2.42$_{-0.43}^{+0.43}$  \\ %0.89            int7
int6 & 224 & 234 & -0.34$_{-0.24}^{+0.25}$ & 4.4$_{-1.6}^{+3.1}$ & -1.53$_{-0.11}^{+0.15}$ & 289$_{-135}^{+750}$   & 146.9/142=1.03 &  -1.00$_{-0.10}^{+0.09}$ & -1.72$_{-0.08}^{+0.08}$ & -1.98$_{-0.63}^{+0.38}$  \\ %0.85            int8
int7 & 234 & 284 & -0.61$_{-0.14}^{+0.16}$ & 4.7$_{-1.3}^{+2.3}$ & -1.65$_{-0.13}^{+0.07}$ & 342$_{-156}^{+958}$   & 248.8/233=1.07 &  -1.24$_{-0.06}^{+0.06}$ & -1.78$_{-0.08}^{+0.08}$ & -2.00$_{-1.16}^{+0.55}$  \\ %1.15            int9-13
intA & 164 & 254 & -0.32$_{-0.10}^{+0.10}$ & 4.7$_{-0.8}^{+1.1}$ & -1.52$_{-0.07}^{+0.10}$ & 346$_{-115}^{+229}$ & 430.9/437=0.99 &     -1.00$_{-0.04}^{+0.03}$ & -1.68$_{-0.04}^{+0.04}$ & -2.28$_{-0.24}^{+0.19}$ \\ %0.94            intB
intB & 164 & 314 & -0.50$_{-0.08}^{+0.09}$ & 5.0$_{-0.8}^{+1.1}$ & -1.54$_{-0.09}^{+0.10}$ & 333$_{-118}^{+265}$ & 529.8/503=1.05 &     -1.12$_{-0.03}^{+0.03}$ & -1.71$_{-0.04}^{+0.04}$ & -2.27$_{-0.27}^{+0.22}$ \\ %0.96            intA
\enddata
\end{deluxetable}
\clearpage

\begin{figure}[!]
%\begin{figure}[!hbp]
\centering
   \includegraphics[width=.49\textwidth, angle=0]{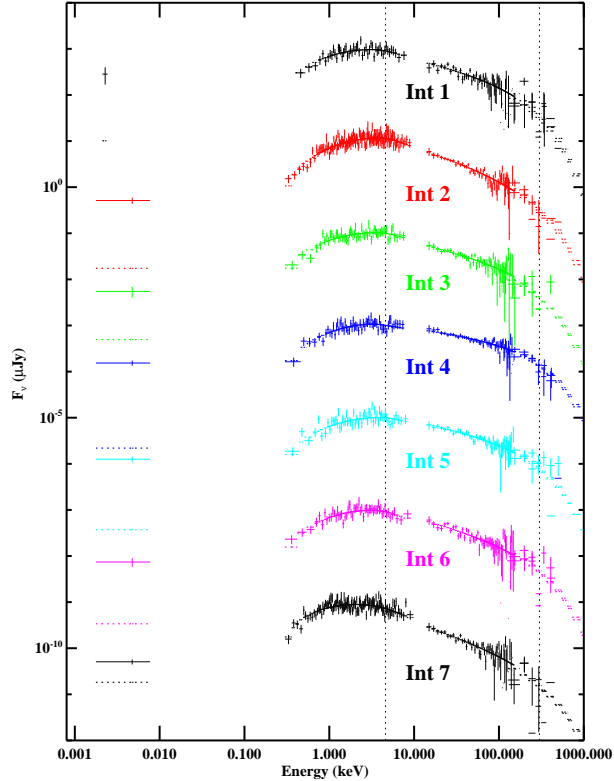}
   \caption{Best fit prompt spectra of GRB 110205A, different colors
   present different intervals (see
   Figure \ref{fig:110205A_prompt_lc} for interval definition). Note that WAM data above 400 keV is not
   shown in the figure due to large uncertainities when transforming counts to flux density in Xspec
   above this energy range. Intervals 2 to 7 are
   shifted by a factor of 10$^{-2}$ accumulatively for a purpose of
   clarity. UVOT optical data is not included during the fitting, but it
   is shown in the best fit spectrum. Solid lines present the observed data
   while dashed lines present the prediction of best fit $bandcut$ model.
   The two vertical lines show the mean value of E$_0$ and E$_1$ from the
   fitting.\label{fig:prompt_sed}}
\end{figure}

Figure \ref{fig:prompt_spec_par} shows the evolution of different
parameters from the time-resolved spectral fitting between 146 s and
284 s. The two break energies, $E_0$ and $E_1$, remain almost
constant during this time range, with $E_0 \sim 5$ keV and $E_1 \sim
300$ keV. However, the large errors for $E_0$ and $E_1$ prevent us
from drawing a firmer conclusion regarding the temporal evolution of
the two break energies.
The low energy photon index, $\alpha$, also shows no statistically
significant evolution. However, the high energy photon
index, $\beta$, does show a weak evolution: it becomes slightly harder
during the brightest BAT peak around 210 s. Over all, the
$\alpha$ value during the time range is around -0.35 and $\beta$ is
around -1.5.
The peak energy derived from interval B (intB) is E$_p$ = (2 + $\beta$)E$_1$
= 153$^{+121}_{-54}$ keV.

Although the UVOT optical data are not included in the spectral
fitting, they are shown in the best fit spectra in Figure
\ref{fig:prompt_sed}. For all the intervals, the optical data are
above the extrapolation of the best spectral fits for high energies.
A speculation may be that the observed $\alpha$ is somehow harder
than the expected value of the synchrotron model. However, even if
we set $\alpha$ to the predicted value, -2/3, the optical
data are still above the best fit spectra in some intervals.
This suggests that the optical emission may have a different
origin from the high energy emission (see \S\ref{sec:models}
for more detailed discussion).

\subsection{Afterglow analysis}
\subsubsection{Light curves}

%\begin{figure}[!]
\begin{figure}[!hbp]
\centering
   \includegraphics[width=.49\textwidth]{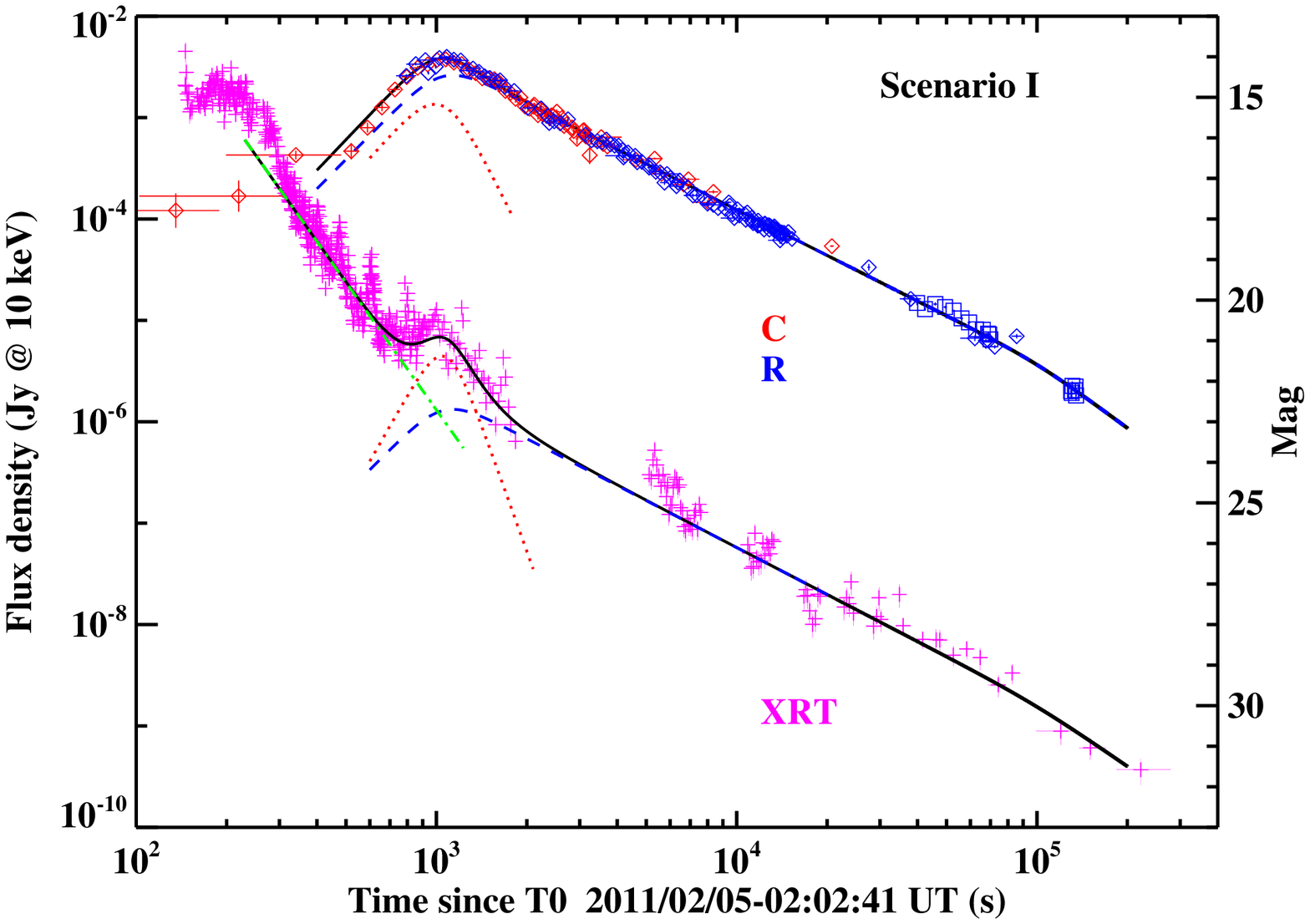}
   \includegraphics[width=.49\textwidth]{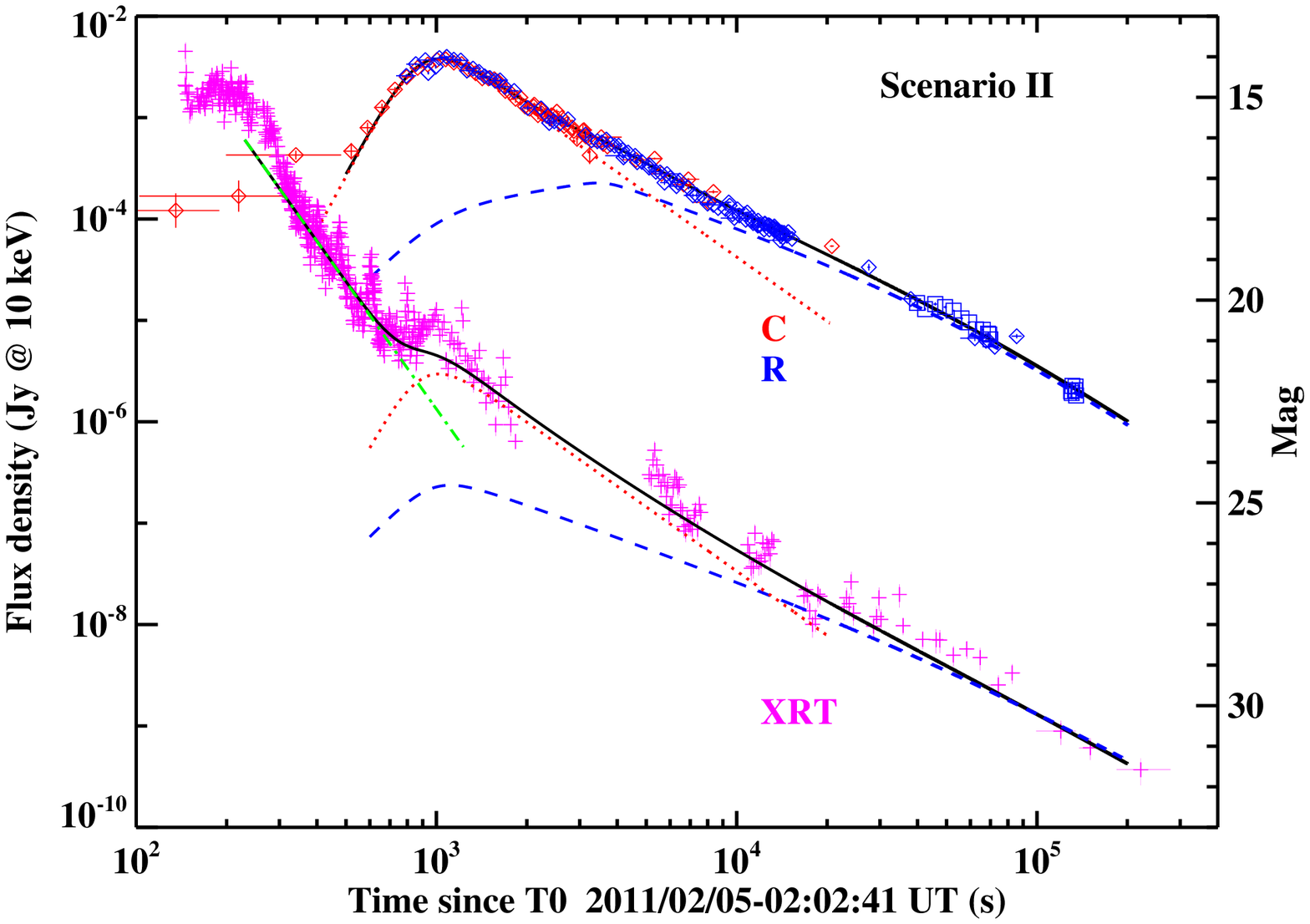}
   \caption{GRB 110205A light curves from XRT data ($pink$ cross) and optical data,
   including  $C$ band ($red$ diamond) and $R$ band ($blue$ square). Two scenarios are
   considered. Both scenarios consider superposition
   of FS (dashed line) and RS (dotted line) contribution. The main difference is that Scenario I
   (left panel) considers the optical peak is dominated by both FS and RS combination
   while Scenario II considers the optical peak is dominated by RS only.
   The dot-dashed line represents the steep decay phase
   for X-ray data. The solid line represents the combination of all components.\label{fig:XRT_Opt_lc}}
\end{figure}

\begin{figure}[!hbp]
\centering
   \includegraphics[width=.49\textwidth, angle=0]{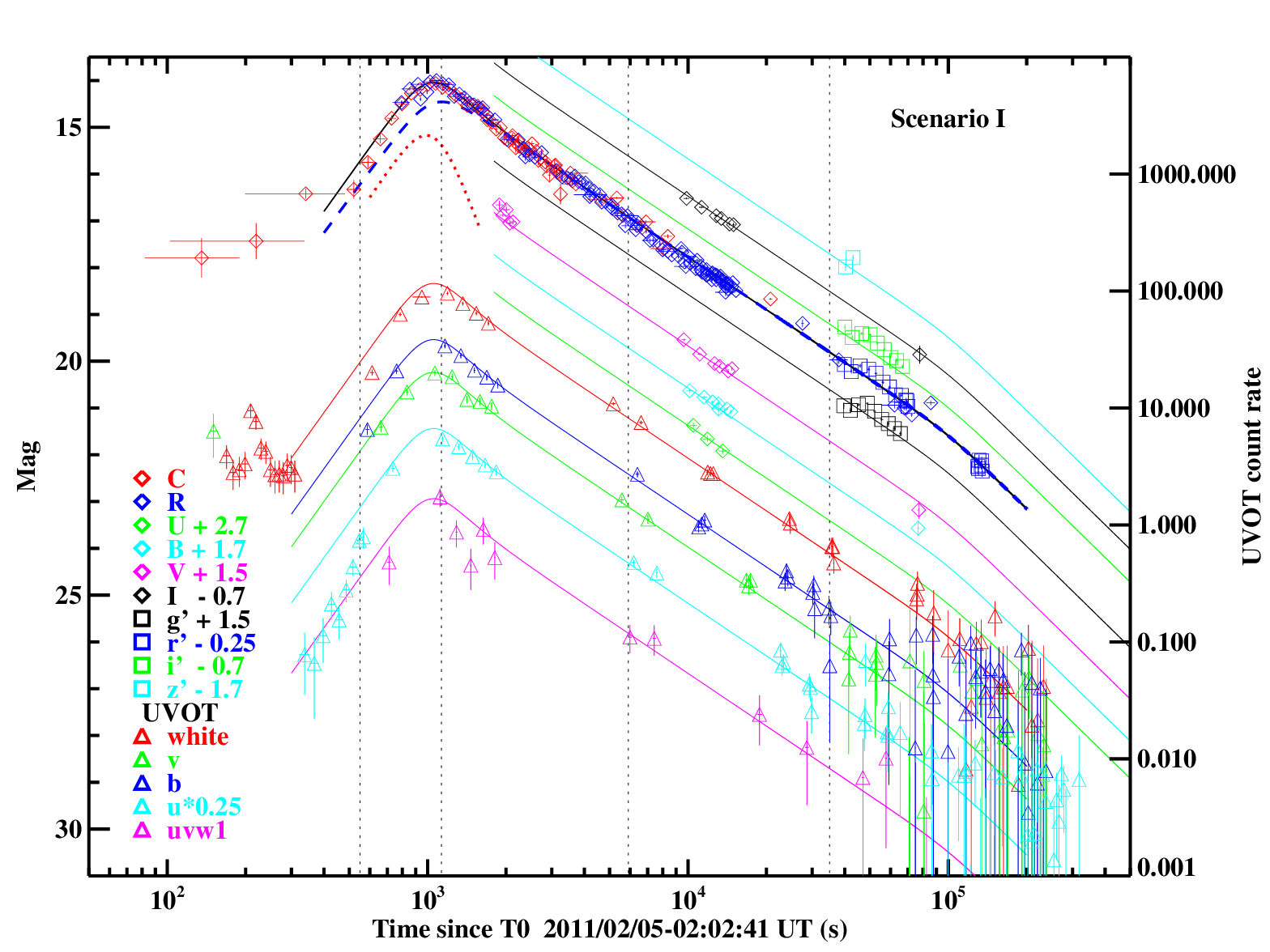}
   \includegraphics[width=.49\textwidth, angle=0]{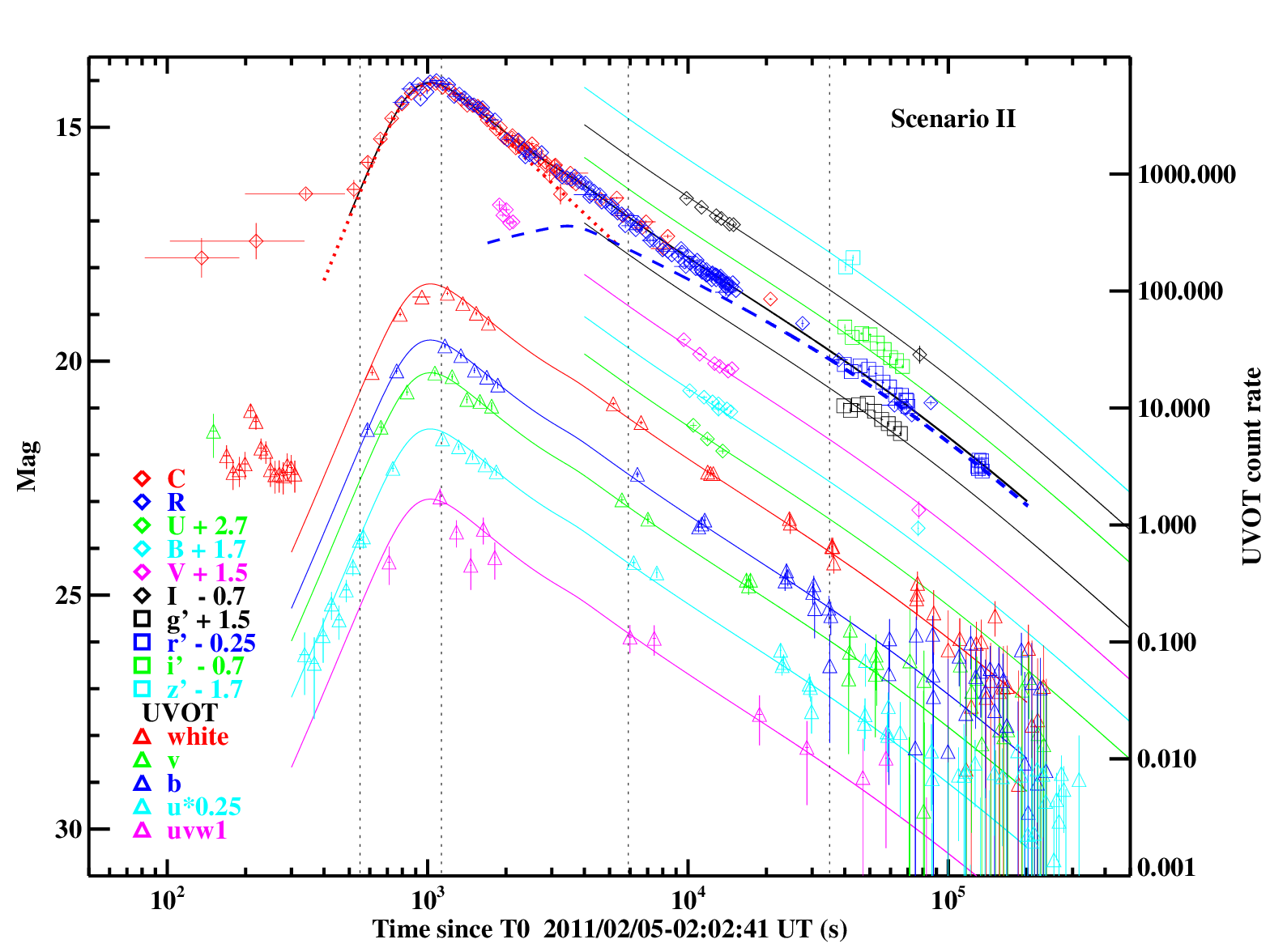}
   \caption{Optical light curves of GRB 110205A re-scaled from $R$ band
   best fit. The early fast rise behavior in optical band can be
   better explained by Scenario II (right panel) than Scenario I (left
   panel). \label{fig:Opt_lc}}
\end{figure}

Shortly after the prompt emission, the optical light curve is
characterized by a surprisingly steep rise and a bright peak
around 1100 s. The early steep rise, starting around 350 s, is
observed by both ROTSE and UVOT and the peak, which is wide and
smooth, reaches $R$ = 14.0 mag at 1073 s after the burst. Such a bright
peak around this time with such a steep rise is rare and unusual
(Oates et al. 2009; Panaitescu \& Vestrand 2011) and has only been reported
in a few cases (e.g. Volnova et al. 2010; Nardini et al. 2011). A
peak brightness of $R$ = 14.0 mag at $\sim$ 1100 s after the burst
ranks the optical afterglow as one of the brightest ever
observed in this same time range (Akerlof \& Swan 2007).
Following the peak, the light curve decays monotonically, as shown in Figure \ref{fig:XRT_Opt_lc},
and displays a slight flattening around 3000 s. Around $5\times 10^4$ s, there is
a re-brightening feature observed in the $g',r',i',z'$ bands by LOT
(Figure \ref{fig:Opt_lc}) and a final steepening is observed after $\sim 10^5$s.
Overall, no substantial color evolution is observed in the optical band.

The X-ray data show a different temporal behavior (see also Figure
\ref{fig:XRT_Opt_lc}). Shortly after the prompt emission, the
light curve has a very steep decay, between 350 s
and 700 s. This is followed by a small re-brightening bump around 1100 s,
the peak time of the optical light curve, and a monotonic decay afterwards.
There might be a late X-ray flare around 5000 s, but the lack of X-ray data just before $\sim$ 5000 s 
prevents any robust conclusion from being drawn.

The light curves were fit with one or the superposition of two
broken-power-law functions. The broken power-law function
has been widely adopted to fit afterglow light curves with both the rising
and decay phases (e.g. Rykoff et al. 2009) and works well for most cases
(e.g. Liang et al. 2010). The function can be represented as:
\begin{equation}
f = \left(\frac{t}{t_b}\right)^{\alpha_1} \Big{[} 1 +
\left(\frac{t}{t_b}\right)^{s({\alpha}1-{\alpha}2)}\Big{]}^{-\frac{1}{s}},
\end{equation}
where $f$ is the flux, $t_b$ is the break time, ${\alpha}1$ and
${\alpha}2$ are the two power law indices before and after the
break, and $s$ is a smoothing parameter.
According to this definition, the peak time $t_p$, where the flux reaches
maximum, is
\begin{equation}
t_p = t_b \left(\frac{\alpha1}{-\alpha2}\right)^{\frac{1}{s(\alpha1
- \alpha2)}}.
\end{equation}
If a multi broken-power-law function is required to present more than one break times,
equation 2 in van Eerten \& MacFadyen (2011) is adopted.
We first tried one broken power law component, and found that it could not fully represent
the feature near the optical peak, mainly because of its unusually late and steep
rising feature, which was not observed in previous GRBs,
and a slight flattening feature around 3000 s after the peak shown in $R$ band.
Noticing that there is a peak both in the
optical and the X-ray light curves around 1100 s and that
the optical light curve flattens around 3000 s,
we speculate that there is a significant contribution of emission
from the reverse shock (RS). A RS contribution to the X-ray
band has not been well identified in the past. Theoretically, the RS
synchrotron emission peaks around the optical band so that
its synchrotron extension to the X-ray band is expected to be weak.
In any case, under certain conditions, it is possible that the RS
synchotron (Fan \& Wei 2005; Zou et al. 2005) or SSC (Kobayashi et al.
2007) emission would contribute significantly to the X-ray band
to create a bump feature. In order to account for both the FS and
the RS components, we fit both the optical and X-ray light curves with
the superposition of two broken-power-law functions.
For the X-ray light curve, an additional single power-law component
was applied for the steep decay phase, as usually seen in $Swift$
afterglows (e.g. Tagliaferri et al. 2005).

Two light curve models have been adopted. The standard afterglow model predicts that
the blast wave enters the deceleration phase as the reverse
shock crosses the ejecta (Sari \& Piran 1995; Zhang et al. 2003).
If the FS $\nu_m$ is already below the optical band at the crossing
time, both the FS and the RS would have a same peak time ($t_p$)
in the optical band. This defines our first scenario, in which the optical peak is generated by both the RS and the FS.
If, however, $\nu_m$ is initially above the
optical band at the deceleration time, the FS optical light curve
would display a rising feature ($\propto t^{1/2}$, Sari et al. 1998)
initially, until reaching a peak at a later time when $\nu_m$
crosses the optical band. This is our second scenario. In this
case, the early optical peak is mostly dominated by the RS component.
This is the Type I light curve identified in Zhang et al. (2003)
and Jin \& Fan (2007). We now discuss the two scenarios in turn.

Scenario I : We performed a simultaneous fit to both optical $R$ band data
and X-ray data by setting the same break time in the two bands. A late time break is
invoked to fit both the $R$ band and X-ray light curves. However, we
exclude the re-brightening feature around 5$\times$10$^4$ s in the optical
band, and the steep decay phase of X-rays before 400 s, which is likely
the tail of the prompt emission (Zhang et al. 2006).
For the FS component, the rising temporal index is fixed to be 3 based
on the slow cooling ISM model during the pre-deceleration 
phase\footnote{In the coasting phase of a thin shell decelerated by a constant
density medium, the forward shock have the scalings $\nu_m \propto t^0$,
$\nu_c \propto t^{-2}$, and $F_{\nu,m} \propto t^3$. For $\nu_m < \nu_{opt}
< \nu_c$, one has $F_\nu \propto t^3$.}
(e.g. Xue et al. 2009; Shen \& Matzner 2012 ). 
The rising slope of the RS component is left
as a free parameter to be constrained from the data. The best simultaneous fit results
are summarized in Table \ref{tab:fit_res_R_X_lc} and shown in Figure
\ref{fig:XRT_Opt_lc} (left panel). The $red$ dot, and $blue$ dashed lines
represent the RS and FS components, respectively.
The $black$ solid line is the sum of the all components.

\begin{deluxetable}{l|cc|cc}
 \tablewidth{0pt}
 \tablecaption{~ Best fit result from $R$ band and X-ray light curves.\label{tab:fit_res_R_X_lc}}
  \tablehead{\colhead{} & \colhead{Scenario I} & \colhead{} &  \colhead{Scenario II} & \colhead{}}
\startdata
Par & Optical & X-ray & Optical & X-ray\\
\hline
FS & & & FS & \\
\hline
 $\alpha_{f1}$  & 3$^*$ & 3$^*$   & 3$^*\to$0.5$^*$ & 3$^*$ \\
 t$_p$ (s)          & 1064$\pm$42 & 1064$^{\#}$    & 3.64$\pm$1.0 $\times$10$^{3}$ & 1021$\pm$26\\
 F$_p$ (Jy)         & 4.96$\times$10$^{-3}$ & 1.32 $\times$10$^{-6}$ & 4.35$\times$10$^{-4}$ & 1.42$\times$10$^{-7}$ \\
                    & 14.48 mag             &                        & 17.12 mag             &                       \\
 $\alpha_{f2}$  & -1.50$\pm$0.04 & -1.54$\pm$0.10  & -1.01$\pm$0.01 & -1.00$^*$ \\
 t$_{jb}$ (s)       & 1.0$\pm$0.2 $\times$10$^5$ & 1.0$\times$10$^5$$^\#$ & 5.44$\pm$0.2$\times$10$^4$ & 5.44$\times$10$^{4\#}$ \\
 $\alpha_{f3}$  & -2.18$\pm$0.8 & -2.05$\pm$0.5  & -2.05$\pm$0.7 & -1.75$\pm$1.0 \\
 & & & &\\
RS & & & RS & \\
\hline
 $\alpha_{r1}$  & 3.32$\pm$1.2 & 5.19$\pm$1.3  & 5.5$^*$ & 5.5$^*$ \\
 t$_{p}$(s)        & 1064$^\#$ & 1064$^\#$    & 1021$^\#$ & 1021$^\#$ \\
 F$_p$ (Jy)        & 2.47$\times$10$^{-3}$ &  4.40 $\times$10$^{-6}$  & 7.19$\times$10$^{-3}$ & 3.59$\times$10$^{-6}$ \\
                   & 15.24 mag             &                          & 14.07 mag             &                       \\
 $\alpha_{r2}$  & -5.90$\pm$1.0 & -8.26$\pm$1.3  & -2.10$^*$ & -2.10$^*$ \\
 & & & &\\
 & & Steep decay &  & Steep decay \\
 \hline
 $\alpha_{sd}$  &   &   -4.18$\pm$0.2 &   &   -4.16$\pm$0.2 \\
\enddata
\tablenotetext{*}{indicates the parameter is fixed during the fitting}
\tablenotetext{\#}{indicates the parameter is simultaneously fit for both optical and X-ray data}
\end{deluxetable}

Next we apply the same best $R$-band fit parameters and re-scale to the optical
data in other bands.
The results are shown in Figure \ref{fig:Opt_lc} (left panel).
Light curves in different bands are properly shifted for 
clarity. As we can see, the model quite adequately describes
most light curves.
The only exception is the UVOT-$u$ band,
in which the model curve overpredicts the early time flux
between 300 s to 600 s.

%\begin{figure}[!]

Scenario II : In this scenario, the bright peak around 1100 s and
the steep rise phase at early times in the optical band are contributed
by the RS only. The FS component shows up later.
According to the afterglow model, the FS component is characterized
by a double broken power law with a rising index +3 before the deceleration
time ($t_p$ of the RS), +0.5 before the FS peak, and a normal decay
(decay index to be fit from the data) after the FS peak (e.g. Sari
et al. 1998; Zhang et al. 2003; Xue et al. 2009).
For the RS component, the rising index is fixed to be 5.5, which is the mean
value of a single power law fit to the $R$ band data ($\alpha \sim$ 5) and the UVOT-$u$
band data ($\alpha \sim$ 6) only.
In the X-ray band, we still use a similar model as Scenario I with
a superposition of a RS and a FS component. The model parameters are
not well constrained,
especially for the X-ray peak around 1100 s due to its narrow peak. We tried
to fix several parameters in order to reach an acceptable fitting
for this scenario.
The best simultaneous fit results for both $R$ band
 and X-ray band data
are also summarized in Table \ref{tab:fit_res_R_X_lc},
and shown in Figure
\ref{fig:XRT_Opt_lc} (right panel). Similar to Scenario I, we then
used the same model and parameters derived from the $R$ band fit re-scale to
other light curves. The results are shown
in Figure \ref{fig:Opt_lc} (right panel), again with shifting.
This scenario can well
explain the early fast rise behavior in all bands.

Comparing the two scenarios, we find that Scenario I can represent
most optical data, and can better account for the X-ray data than
Scenario II. However, it can not well match the early very steep
rise in the UVOT-$u$ band. On the other hand, Scenario II,
can represent the early fast rising behavior in all the
optical band (including UVOT-$u$ band), but the fits to the X-ray
data are not as good as Scenario I.
We note that our scenario II is similar to another variant of
scenario II recently proposed by Gao (2011), but we conclude that
the optical data before $\sim$ 350 s are generated by the optical
prompt emission.

\subsubsection{Afterglow SED analysis}
In order to study the spectral energy distribution (SED) of the
afterglow, we selected 4 epochs when we have the best multi-band
data coverage: 550 s, 1.1 ks, 5.9 ks and 35 ks. Since no significant color
evolution is observed in the optical data, we interpolate (or extrapolate if necessary) the optical
band light curve to these epochs when no direct observations are
available at the epochs. The optical data used for
constructing the SED are listed in Table \ref{tab:sed_data}. The X-ray
data are re-binned using the data around the 4 epochs. The spectral
fitting is then performed using Xspec software. During the fitting,
the Galactic hydrogen column density, N$_H$, is fixed to
1.6$\times$ 10$^{20}$ cm$^{-2}$ (Kalberla et al. 2005), and the host galaxy hydrogen column
density is fixed to 4.0$\times$ 10$^{21}$ cm$^{-2}$. These are
derived from an average spectral fitting of the late time XRT PC
data. We tried both the broken-power-law and the single power-law models
for the afterglow SED. For the broken-power-law model, we set $\Gamma_2$ =
$\Gamma_1$ - 0.5, assuming a cooling break between the optical and
X-ray bands. Meanwhile, we also investigated the use of three
different extinction laws, namely
the Milky Way (MW), Large and Small Magellanic Clouds (LMC and SMC),
for the host galaxy extinction model. The E(B-V) from the
Galactic extinction is set to 0.01 (Schlegel et al. 1998) during the fitting. The best
fit results are listed in Table \ref{tab:sed_ref}.

\begin{deluxetable}{ccccc}
 \tablewidth{0pt}
 \tablecaption{~ SED data  (in magnitude) at 4 different epoch\label{tab:sed_data}}
  \tablehead{\colhead{time} & \colhead{550 s} & \colhead{1.13 ks} & \colhead{5.9 ks} & \colhead{35 ks} }
\startdata
% - & - & - & - &  - \\
uvw1 & 20.13$\pm$0.50 & 16.91$\pm$0.18 & 19.71$\pm$0.30 & 22.50$\pm$0.70  \\
u & 17.13$\pm$0.20 & 14.98$\pm$0.06 & 17.50$\pm$0.10 & 20.47$\pm$0.30  \\
b &  17.54$\pm$0.16 & 15.23$\pm$0.05 & 17.92$\pm$0.07 & 20.97$\pm$0.30  \\
v &  17.13$\pm$0.30 & 14.68$\pm$0.06 & 17.45$\pm$0.14 &  20.63$\pm$0.50 \\
white &  17.80$\pm$0.12 & 15.25$\pm$0.05 & 17.92$\pm$0.05 &  20.74$\pm$0.22 \\
\\
U & - & - & 17.74$\pm$0.1 & 20.69$\pm$0.1  \\
B & - & - & 18.05$\pm$0.1 & 20.86$\pm$0.1  \\
V & - & - & 17.24$\pm$0.1 & 20.20$\pm$0.1  \\
R & 16.07$\pm$0.20 & 14.10$\pm$0.06 & 16.84$\pm$0.1 & 19.73$\pm$0.1  \\
I & - & - & 16.39$\pm$0.1 & 19.21$\pm$0.1  \\
\\
g & - & - & - &  20.76$\pm$0.07 \\
r & - & - & - &  20.12$\pm$0.05 \\
i & - & - & - &  19.76$\pm$0.06 \\
z & - & - & - &  19.46$\pm$0.09 \\

\enddata
\end{deluxetable}

%\begin{figure}[!]
\begin{figure}[!hbp]
\centering
   \includegraphics[width=.49\textwidth]{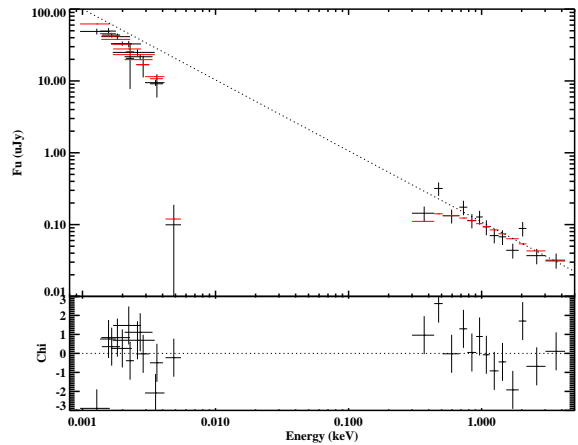}
   \caption{Afterglow SED of GRB 110205A at 35 ks. $Black$ points are the
   observed data, $red$ points are the predicted value from the model.
   Dotted line is the original power-law model without absorption or
   extinction.\label{fig:sed_plot}}
\end{figure}

The 5.9 ks and 35 ks SEDs have the best data coverage, and
they are ascribable to the FS component only. We therefore use
these two SEDs to constrain the extinction properties of the afterglow.
We find that the SMC and LMC dust models provide acceptable and better
fits than MW dust model. The data are equally well fit by the broken-power-law
and the single power-law model. For the broken-power-law model, the
break energy is found to be within the optical band (0.0025 $\pm$
0.0006 keV). Within the 3-$\sigma$ error, one cannot separate the optical
and X-ray data to two different spectral regimes. The lack of clear breaks
in optical light curves between 5.9 and 35 ks also disfavors the possibility of the
break energy passing through the optical band.

\subsection{Host galaxy search}
%\begin{figure}[!]
\begin{figure}[!hbp]
\centering
   \includegraphics[width=.49\textwidth]{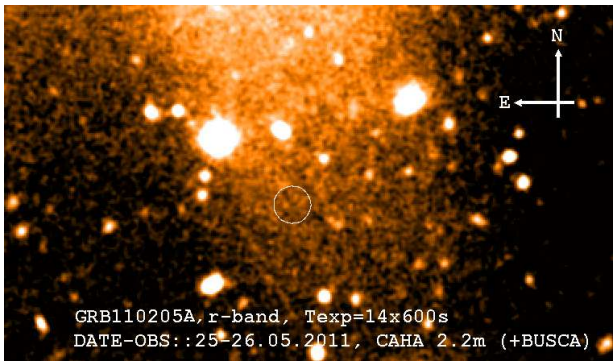}
   \caption{Host galaxy searching from 2.2m Calar Alto telescope taken in $r'$ band.
   The center of the circle indicates the afterglow location. No clear source
   is detected down to $r'$ $\sim$ 24.8 within 5" of the afterglow location.\label{fig:host}}
\end{figure}

We have performed a deep search for the host galaxy of the GRB.
Observations were performed with the 2.2 m Calar Alto telescope 3.7
months after the burst. Images were taken with the BUSCA instrument in the $u',g',r'$ and $i'$
bands under good seeing condition, with image resolution of $\sim$ 0.9". Co-adding was
applied to a set of individual images in order to obtain a deeper limiting
magnitude. Figure \ref{fig:host} shows one of the images taken in the $r'$
band. The center of the circle indicates the afterglow location.

No clear source was detected near the afterglow location within a radius
of 5".
The typical 3$\sigma$ upper limits (AB magnitudes) are: $u'\sim$ 24.1,
$g'\sim$ 24.4, $r'\sim$ 24.8 and $i'\sim$ 25.2. A non-detection of the GRB host
galaxy at $r'\sim$ 24.8 is not surprising since a lot of GRB host galaxies
are faint (e.g. Savaglio et al. 2009) or not detected at all (e.g. Ovaldsen
et al. 2007). It is also superseded by the much deeper observation reported by Cucchiara et al. (2011)
down to upper limit of $r'\sim$ 27.21 magnitudes.
At redshift $z$=2.22, a magnitude of $r'\sim$ 27.2 corresponds
to an absolute magnitude M$_{r'}$ $\sim$ -19.1, the upper limit is fainter than 70\% of the GRB host galaxies
compared with large host galaxies samples searched systematically by some groups (see Figure 3. in Jakobsson et al.
2010; Pozanenko et al. 2008).

\begin{deluxetable}{cccccccc}
 \tablewidth{0pt}
 \tabletypesize{\scriptsize}
 \tablecaption{~ SED fitting at different epoch\label{tab:sed_ref}}
  \tablehead{\colhead{Model} & \colhead{time} & \colhead{E(B-V)} & \colhead{$\Gamma_1$} & \colhead{E$_b$ (keV)} & \colhead{$\Gamma_2$=$\Gamma_1$+0.5} & \colhead{Norm} & \colhead{$\chi^2$/dof} }
\startdata
MW  & Bknplaw &  &   &   &   &  & \\
\\
    & 550 s   &  0.11$_{-0.02}^{+0.02}$ & 1.36$_{-0.034}^{+0.033}$ & 0.41$_{-0.107}^{+0.074}$ & 1.86 &    0.23$_{-0.019}^{+0.039}$ & 124.8/81=1.54  \\
    & 1.13 ks &  0.19$_{-0.004}^{+0.009}$ & 2.10$_{-0.022}^{+0.021}$ & 4.94$_{-4.93}^{+4.93}$ & 2.60 &    0.038$_{-4.0E-3}^{+4.39E-3}$ & 13.6/4=3.6 \\
    & 5.9 ks  &  0.16$_{-0.007}^{+0.02}$ & 1.59$_{-0.027}^{+0.0100}$ & 1.00E-03$_{-1.0E-3}^{+1.5E-3}$ & 2.09 &    0.086$_{-0.022}^{+4.34E-3}$ & 20.2/23=0.88 \\
    & 35 ks   &  0.13$_{-0.008}^{+0.009}$ & 1.57$_{-0.012}^{+0.011}$ & 1.70E-03$_{-1.57E-4}^{+3.66E-3}$ & 2.07 &   4.03E-03$_{-1.89E-3}^{+2.15E-4}$ & 38.9/22=1.77 \\
\\
MW  &  plaw   &  &   &   &   &  & \\
\\
    & 550 s   &  0.26&                  - & - & 1.70                           &   0.14       & 179.8/82= 2.19 \\
    & 1.13 ks &  0.19$_{-0.009}^{+0.009}$ & - & - & 2.10$_{-0.022}^{+0.022}$ & 0.038$_{-0.004}^{+0.004}$ & 13.6/5=2.72 \\
    & 5.9 ks  &  0.16$_{-0.019}^{+0.020}$ & - & - & 2.09$_{-0.027}^{+0.027}$ & 0.027$_{-1E-4}^{+1E-4}$ & 20.2/24= 0.84  \\
    & 35 ks   &  0.13                        & - & - & 2.07             & 1.59E-04      & 55.6/23= 2.42 \\
\hline \\
LMC & Bknplaw &  &   &   &   &  & \\
\\
    & 550 s   &  0.10$_{-0.017}^{+0.021}$ & 1.39$_{-0.033}^{+0.033}$ & 0.36$_{-0.151}^{+0.0951}$ & 1.89 &    0.25$_{-0.027}^{+0.076}$ & 116.1/81= 1.43  \\
    & 1.13 ks &  0.16$_{-0.004}^{+0.013}$ & 1.64$_{-0.030}^{+0.013}$ & 2.88$_{3.34E-3}^{+2.60E-3}$ & 2.14 &    0.72$_{-0.027}^{+0.035}$ & 8.5/4=2.13 \\
    & 5.9 ks  &  0.13$_{-0.019}^{+0.018}$ & 1.61$_{-0.049}^{+0.099}$ & 2.87E-03$_{-9.33E-4}^{+7.01E-4}$ & 2.11 &    0.051$_{-0.020}^{+0.007}$ & 16.1/23= 0.70  \\
    & 35 ks   &  0.10$_{-0.007}^{+0.006}$ & 1.55$_{-0.014}^{+0.016}$ & 2.07E-03$_{-3.48E-4}^{+5.14E-4}$ & 2.05 &   3.70E-03$_{-3.85E-4}^{+2.58E-4}$ & 26.4/22= 1.20 \\
\\
LMC &    plaw &  &   &   &   &  & \\
\\
    & 550 s   &  0.24$_{-0.021}^{+0.023}$ & - & - & 1.75$_{-0.031}^{+0.033}$ & 0.141$_{-3.85E-3}^{+3.93E-3}$ & 162.5/82= 1.98 \\
    & 1.13 ks &  0.14$_{-0.006}^{+0.006}$ & - & - & 2.09$_{-0.021}^{+0.021}$ & 0.039$_{-4.13E-3}^{+4.45E-3}$ & 13.9/5=2.78 \\
    & 5.9 ks  &  0.10$_{-0.011}^{+0.011}$ & - & - & 2.04$_{-0.020}^{+0.020}$ & 2.71E-03$_{-1.35E-4}^{+0.001}$ & 19.4/24= 0.81 \\
    & 35 ks   &  0.07$_{-0.009}^{+0.006}$ & - & - & 2.01$_{-0.015}^{+0.015}$ & 1.65E-04$_{-1.05E-05}^{+1.04E-05}$ & 39.0/23= 1.70  \\
\hline \\
SMC & Bknplaw &  &   &   &   &  & \\
\\
    & 550 s   &  0.08$_{-0.014}^{+0.016}$ & 1.38$_{-0.032}^{+0.032}$ & 0.37$_{-0.14}^{+0.09}$ & 1.88&    0.25$_{-0.026}^{+0.068}$ & 114.3/81= 1.41 \\
    & 1.13 ks &  0.11$_{-0.005}^{+0.005}$ & 2.06$_{-0.021}^{+0.020}$ & 5.05$_{-4.93}^{+-4.93}$ & 2.56 &    0.039$_{-2.24E-3}^{+4.56E-3}$ & 24.7/4=6.18 \\
    & 5.9 ks  &  0.09$_{-0.004}^{+0.005}$ & 1.56$_{-0.028}^{+0.024}$ & 2.74E-03$_{-5.05E-4}^{+9.4E-4}$ & 2.06 &    0.052$_{-7.14E-3}^{+5.57E-3}$ & 17.3/23= 0.75  \\
    & 35 ks   &  0.08$_{-0.006}^{+0.005}$ & 1.54$_{-0.013}^{+0.016}$ & 2.43E-03$_{-3.7E-4}^{+4.53E-4}$ & 2.04 &    3.42E-03$_{-2.75E-4}^{+2.78E-4}$ & 24.3/22= 1.1 \\
\\
SMC &    plaw &  &   &   &   &  & \\
\\
    & 550 s   &  0.19 &   - & - & 1.75   &  0.136  &  189.5/82= 2.31 \\
    & 1.13 ks &  0.11$_{-0.005}^{+0.005}$ & - & - & 2.06$_{-0.020}^{+0.021}$ & 0.039$_{-4.12E-3}^{+4.40E-3}$ & 24.7/5=4.14 \\
    & 5.9 ks  &  0.07$_{-0.007}^{+0.007}$ & - & - & 2.01$_{-0.017}^{+0.017}$ & 2.68E-03$_{-1.3E-4}^{+1.3E-4}$ & 22.5/24=0.94 \\
    & 35 ks   &  0.05$_{-0.006}^{+0.005}$ & - & - & 1.98$_{-0.013}^{+0.014}$ & 1.65E-04$_{-1.04E-05}^{+1.03E-05}$ & 36.7/23=1.60 \\
\enddata
\end{deluxetable}
\clearpage

\section{Theoretical Modeling}
\label{sec:models}

The high-quality broad band data of GRB 110205A allow us to model both prompt
emission and afterglow within the framework of the standard fireball shock
model, and derive a set of parameters that are often poorly constrained
from other GRB observations.
In the following, we discuss the prompt emission and afterglow modeling in turn.

\subsection{Prompt emission modeling\label{sec:modeling_prompt}}
\subsubsection{General consideration}		\label{sec:general-prompt}

The mechanism of GRB prompt emission is poorly known. It depends on the
unknown composition of the jet which affects the energy dissipation,
particle acceleration and radiation mechanisms (Zhang 2011). In general,
GRB emission can be due to synchrotron, SSC in the regions where kinetic
or magnetic energies are dissipated, or Compton scattering of thermal
photons from the photosphere. Within the framework of the synchrotron-dominated
model (e.g. the internal shock model, Rees \& M\'esz\'aros
1994; Daigne \& Mochkovitch 1998, or the internal magnetic dissipation
model, e.g. Zhang \& Yan 2011), one can have a broken power law spectrum.
Two cases may be considered according to the relative location of the cooling
frequency $\nu_c$ and the synchrotron injection frequency $\nu_m$:
fast cooling ($\nu_c < \nu_m$) or slow cooling phase
($\nu_m < \nu_c$). The spectral indices are sumarized in Table
\ref{tab:spec_index} (e.g. Sari et al. 1998).

For GRB 110205A, one may connect the two observed spectral breaks
($E_0$ and $E_1$) to $\nu_c$ and $\nu_m$ in the synchrotron model.
Since the spectral index above $E_1$ is not well constrained from
the data,
we focus on the regime below $E_1$.
The expected spectral density ($F_\nu$) power law index
is -0.5 or -(p-1)/2, respectively, for
the fast and slow cooling cases. The observed photon index $\beta
\sim -1.5$ matches the fast cooling prediction closely.
It is also consistent with slow cooling if the electron spectral
index $p=2$. For standard parameters, the prompt emission spectrum
is expected to be in the fast cooling regime (Ghisellini et al.
2000). Slow cooling may be considered if downstream magnetic fields
decay rapidly (Pe'er \& Zhang 2006). The data are consistent with
either possibility, with the fast cooling case favored by the
close match between the predicted value and the data.

\clearpage
\begin{deluxetable}{ccccc}
 \tablewidth{0pt}
 \tablecaption{Spectral indices from synchrotron spectrum prediction\label{tab:spec_index}}
  \tablehead{\colhead{fast cooling} & \colhead{} & \colhead{} & \colhead{} & \colhead{} }
\startdata
 & $\nu < \nu_a$ & $\nu_a < \nu < \nu_c$ & $\nu_c < \nu < \nu_m$ & $\nu_m < \nu $ \\
 & 2 & 1/3 & -1/2 & -p/2 \\
slow cooling & & & & \\
\hline
 & $\nu < \nu_a$ & $\nu_a < \nu < \nu_m$ & $\nu_m < \nu < \nu_c$ & $\nu_c < \nu $ \\
 & 2 & 1/3 & -(p-1)/2 & -p/2 \\
 \\
\hline
Observed mean$^*$ &    & $\sim$0.6 & $\sim$ -0.5 &
\enddata
\tablenotetext{*}{Note: spectral index = $\Gamma$ + 1, i.e., $\alpha$ and $\beta$ in $bandcut$ model + 1.}
\end{deluxetable}
\clearpage

Below $E_0$ (which corresponds to $\nu_c$ for fast cooling or $\nu_m$
for slow cooling), the synchrotron emission model predicts a spectral
index of 1/3. The observed mean value is $\sim$ 0.60, which is harder
than the predicted value. Considering the large errors of the spectral
indices, this is not inconsistent with the synchrotron model.
Furthermore, if the magnetic fields are highly tangled with small
coherence lengths, the emission may be in the ``jitter'' regime.
The expected spectral index can then be in the range of 0 to 1,
consistent with the data (Medvedev 2006).

Overall, we conclude that the observed prompt spectrum is roughly
consistent with the synchrotron emission model in the fast cooling
regime.
This is the first time when a clear two-break spectrum is identified in
the prompt GRB spectrum that is roughly consistent with the prediction
of the standard GRB synchrotron emission model.

The detection of bright prompt optical emission in GRB 110205A
provides new clues to GRB prompt emission physics.
The optical flux density of GRB 110205A is
$\sim$ 20 times above the extrapolation from the best fit X/$\gamma$-ray
spectra. On the other hand, the optical light curve roughly traces that
of $\gamma$-rays. This suggests that the optical emission is related to
high energy emission, but is powered by a different radiation mechanism
or originates from a different emission location.
The case is similar to that of GRB 080319B (Racusin et al. 2008),
but differs from that of GRB 990123 (Akerlof et al. 1999) where the
optical light curve peaks after the main episodes of $\gamma$-ray emission
and is likely powered by the reverse shock (Sari \& Piran 1999;
M\'esz\'aros \& Rees 1999; Zhang et al. 2003; Corsi et al. 2005).

In the following, we discuss several possible interpretations of this 
behavior, i.e. the synchrotron + SSC model 
(Kumar \& Panaitescu 2008; Racusin et al. 2008); 
the internal reverse + forward shock model (Yu, Wang \& Dai 2009); 
the two zone models (Li \& Waxman 2008; Fan et al. 2009); and the
dissipative photosphere models (e.g. Pe'er et al. 2005, 2006;
Giannios 2008; Lazzati et al. 2009, 2011; Lazzati \& Begelman 2010;
Toma et al. 2010; Beloborodov 2010; Vurm et al. 2011).
We conclude that the synchrotron + SSC model and the photosphere model are
disfavored by the data while the other two models are viable interpretations.

\subsubsection{Synchrotron + SSC}

Since the spectral shape of the SSC component is similar to the synchrotron
component (Sari \& Esin 2001), the observed two-break spectrum can be
in principle due to SSC while the optical emission is due to synchrotron.
This scenario is however disfavored since it demands an unreasonably high 
energy budget. The arguments are the following:

We take interval 2 as an example since its flux varies relatively slowly.
Let $h \nu_{opt} \sim 10^{-2.3}$ keV and $h \nu_{\g,p} = E_0 \sim 10^{0.7}$ keV,
the latter being the peak frequency of $F_\nu$ for
the SSC component). Observations suggest that
$F_{\nu_{\g,p}} / F_{\nu, opt} \sim 20$,
(see Figure \ref{fig:prompt_sed} and Table \ref{tab:prompt_spec_par}).
Define $\nu_{p,syn}$ as the synchrotron $F_{\nu}$ peak frequency,
and $\beta_{opt}$ the spectral index around $\nu_{opt}$ ($F_{\nu}
\propto \nu^{\beta_{opt}}$). Then the Compton parameter can be written as
\beq    \label{eq:Y}
Y= \nu_{\g,p} F_{\nu_{\g,p}}/ (\nu_{p,syn} F_{\nu_{p,syn}})
\approx 10^{4.3}~ (\nu_{p,syn}/\nu_{opt})^{-1-\beta_{opt}}.
\eeq
The inverse-Compton (IC) scattering optical depth is $\tau_e \sim
F_{\nu_{\g,p}}/F_{\nu_{p,syn}} \sim Y \nu_{p,syn}/\nu_{\g,p} \sim
10^{1.3}~ (\nu_{p,syn}/\nu_{opt})^{-\beta_{opt}}$.

One constraint ought to be imposed for the SSC scenario, that is -- the
high energy spectrum of the synchrotron component at the lower bound of
the X-ray band, i.e., $\nu_X= 0.3$ keV, must be below the observed flux
density there. Since the spectral indices below and above $\nu_{\g,p}$
are consistent with $1/3$ and $-1/2$, respectively, and the
synchrotron spectral slope above its peak resembles that of its SSC
component, one can express this constraint in terms of
$F_{\nu_{p,syn}} (\nu_X/\nu_{p,syn})^{-1/2} < F_{\nu_{\g,p}}
(\nu_X/\nu_{\g,p})^{1/3}$. 
With numbers plugged in, this translates to a lower limit on
the Compton parameter:
\beq  \label{eq:Y-low}
Y > 10^{2.5} (\nu_{p,syn}/\nu_{opt})^{-1/2}.
\eeq

The inferred high $Y$ (Eq. \ref{eq:Y}) and $\tau_e$ values would
inevitably lead to an additional spectral component due to the 2nd-order IC
scattering (Kobayashi et al. 2007; Piran et al. 2009).
The 2nd IC $F_{\nu}$ spectrum peaks at $h \nu_{ic,2} \sim h \nu_{\g,p}
Y/\tau_e \sim 10^{0.7}~ (\nu_{p,syn}/\nu_{opt})^{-1}$ MeV, with a flux
density $\sim \tau_e F_{\nu_{\g,p}} \sim 10^{1.3}~ F_{\nu_{\g,p}}
(\nu_{p,syn}/\nu_{opt})^{-\beta_{opt}}$. The nice fit of the $bandcut$
model to the XRT-BAT-WAM spectrum rules out a 2nd IC peak below 1 MeV
(see Figure \ref{fig:prompt_sed}), which poses a constraint
$\nu_{p,syn}/\nu_{opt} < 5$.

We then get a constraint 
$Y \gtrsim 10^4$ and $\tau_e \gtrsim 10$! 
This would lead to a serious energy crisis due to the 2nd IC scattering. For
$\nu_{p,syn}/\nu_{opt} \ll 1$, the 2nd IC scattering might be in the
Klein-Nishina regime and then be significantly suppressed, but the
synchrotron peak flux density would be self-absorbed causing the
seed flux insufficient for the 1st IC scattering (Piran et al. 2009). In
conclusion, the SSC scenario is ruled out due to the high $Y$ value inferred.

\subsubsection{Internal reverse-forward shocks}	\label{sec:int_RS_FS}

Next we consider the internal shock model by calculating synchrotron emission
from the reverse shock (RS) and forward shock (FS) due to the collision of two 
discrete cold shells (e.g., Rees \& M\'esz\'aros 1994; Daigne \& Mochkovitch 1998; 
Yu et al. 2009). If the two shells have high density contrast, the synchrotron 
frequencies would peak around the $\gamma$-ray (reverse) and optical (forward)
bands, respectively. 

We first derive the frequency and flux ratio between the two shocks 
(Kumar \& McMahon 2008).
We define shell ``1'' as the fast moving, trailing shell, and shell ``2'' as the slower,
leading shell. We use subscript `$s$' to represent the shocked region.  
The pressure balance at the contact
discontinuity 
gives (e.g., Shen, Kumar \& Piran 2010)
\beq    \label{eq:p-bal}
(\G_{1s}^2-1) n_1 = (\G_{2s}^2-1) n_2,
\eeq
where $\G_{1s}$ and $\G_{2s}$ are the Lorentz factors (LFs) of the unshocked shells, respectively, measured in the shocked region rest frame, and $n_1$, $n_2$ are the unshocked shell densities measured in their own rest frames, respectively. This equation is exact and is valid for both relativistic and sub-relativistic shocks. Using Lorentz transformation of LFs, the above equation can give the shocked region LF $\G_s$ for given $n_1/n_2$, $\G_1$, and $\G_2$. 

\begin{figure}[h]
\includegraphics[width=.487\textwidth, angle=0]{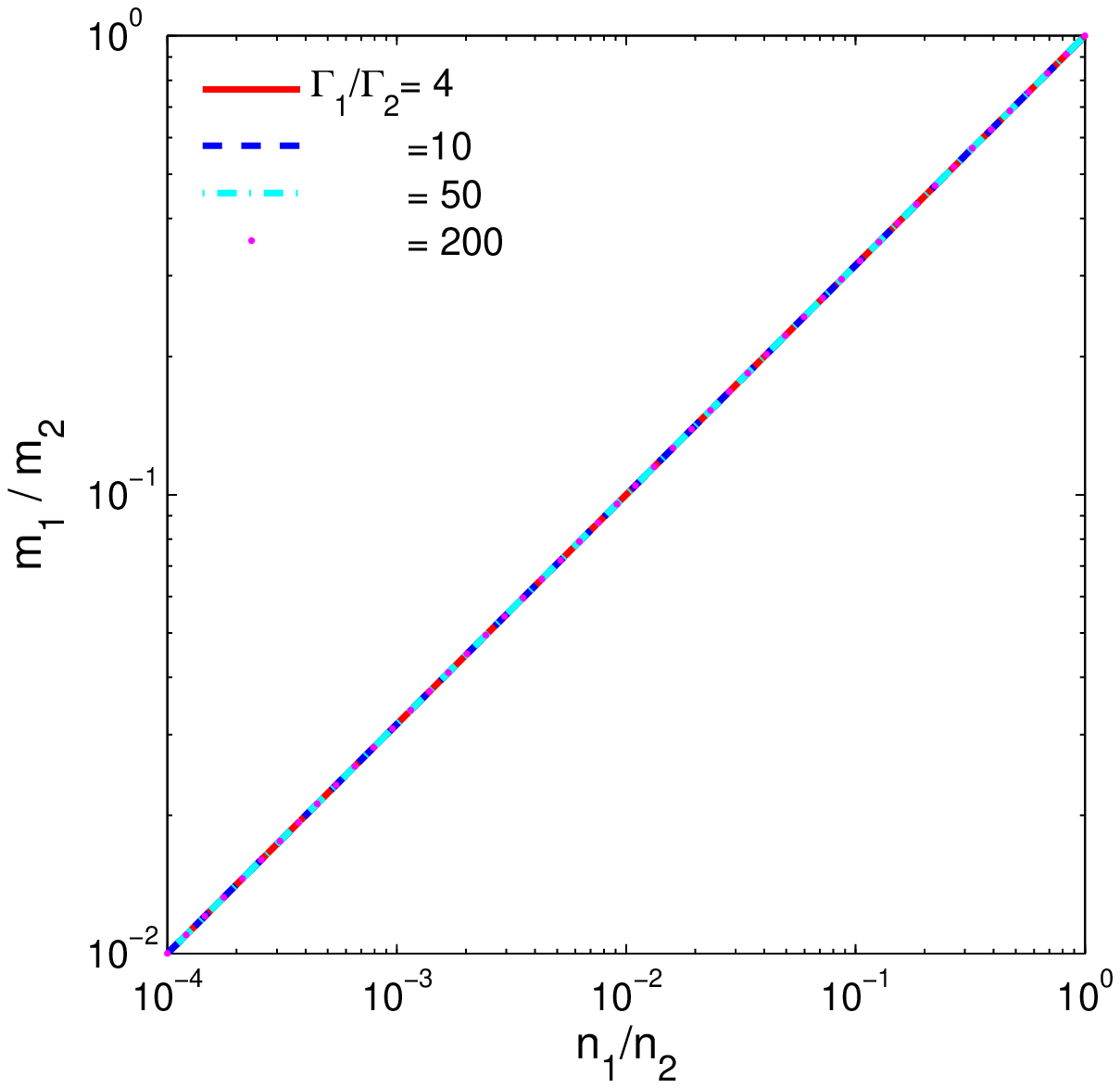}
\includegraphics[width=.487\textwidth, angle=0]{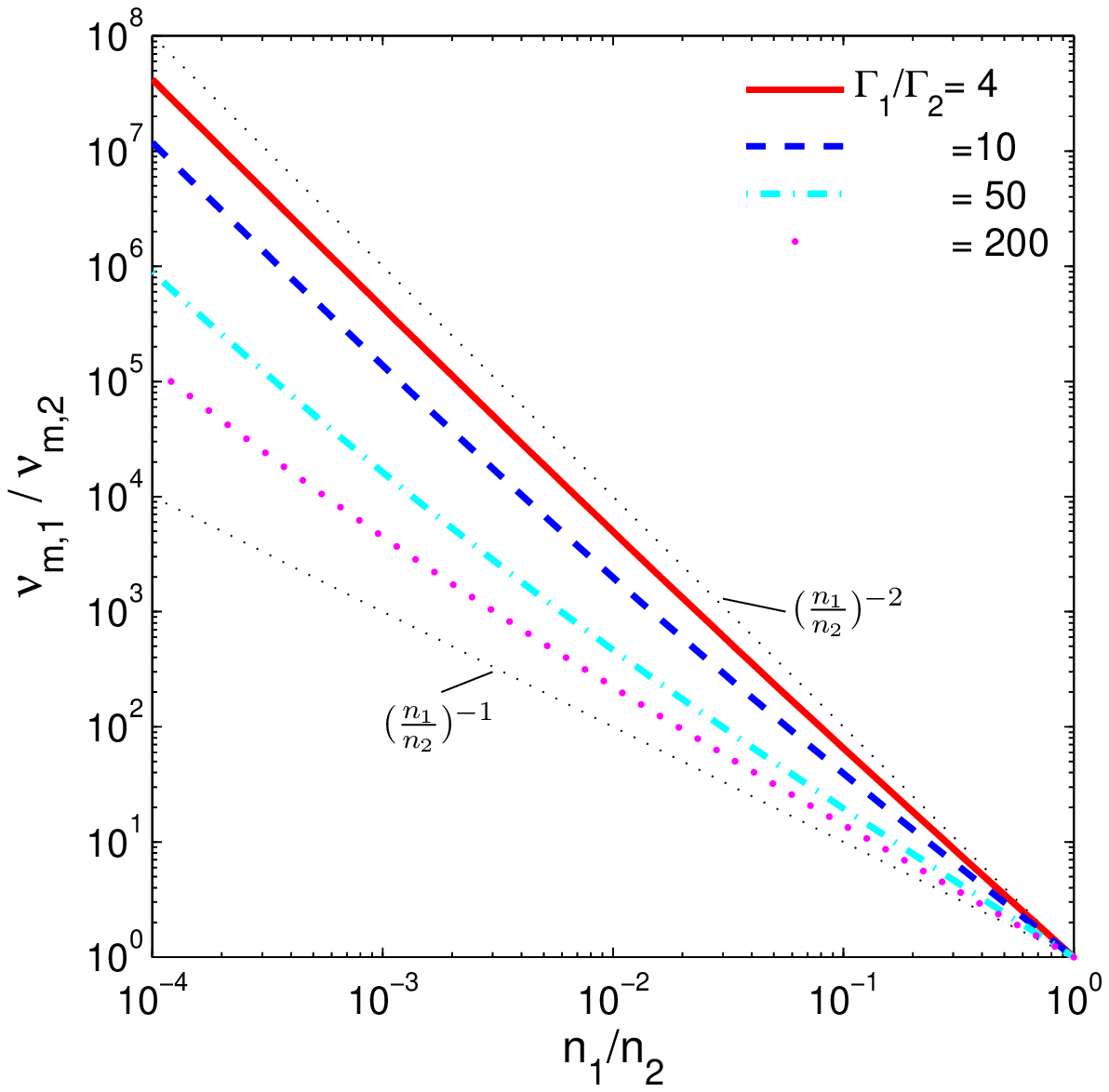}
\caption{The ratios of the shock swept-up masses and synchrotron injection frequencies between the pair of shocks due to collision of two cold relativistic shells. The results are valid for both sub-relativistic and relativistic shocks, and are insensitive to $\G_2$ as long as $\G_2 \gg 1$.  Note that these results can be extrapolated to the region of $n_1 > n_2$, and each result is symmetric about the (1, 1) point.}	\label{fig:rsfs}
\end{figure}

We assume that $\eps_e$, $\eps_B$ and $p$ are the same for both RS and FS. In the shocked region, the magnetic field energy density is $U_B'=B'^2/8\pi= 4\bar{\G}(\bar{\G}-1) \eps_B n$, where $\bar{\G}$ is the relative LF between the downstream and upstream of the shock, which corresponds to $\G_{1s}$ and $\G_{2s}$ for RS and FS, respectively. 
Since the internal energy density is the same in the RS and FS regions (due to pressure balance at contact discontinuity), it is obvious that the two shocked regions have the same $B'$, independent of the strengths of the two shocks. 
The injection energy and the cooling energy of the electrons are $\g_m \propto (\bar{\G}-1)$ and $\g_c \propto 1/(U_B' t')$, respectively. Since synchrotron frequency $\nu \propto \Gamma_s B' \g^2$, 
one finds the frequency ratios to be
\beq    \label{eq:Rnum}
\frac{\nu_{m,1}}{\nu_{m,2}}= \frac{(\G_{1s}-1)^2}{(\G_{2s}-1)^2},
\eeq
\beq    \label{eq:Rnuc}
\frac{\nu_{c,1}}{\nu_{c,2}}= 1.
\eeq
Thus, the injection frequency ratio can be determined for given $\G_1$, $\G_2$ and shell density ratio $n_1/n_2$. We numerically calculate $\nu_{m,1}/\nu_{m,2}$ as a function of $n_1/n_2$ and plot it in Fig. \ref{fig:rsfs}, and find that for different shell LF ratios $\nu_{m,1}/\nu_{m,2}$ lies between $(n_1/n_2)^{-1}$ and $(n_1/n_2)^{-2}$.

The maximum flux density is $F_{\nu,\max} \propto \Gamma_s N_e B'$, where $N_e$ is the total number of the shocked electrons. So one has $F_{\nu,\max,1}/F_{\nu,\max,2}= m_1/m_2$, where $m$ is the shock swept-up mass. We calculate the mass ratio in the rest frame of the shocked region. In this frame the density of the shocked fluid is $4\bar{\G} n$; the rate of mass sweeping is proportional to the sum of the shock front speed and the unshocked fluid speed. We then get
\beq    \label{eq:Rm}
\frac{m_1}{m_2}= \frac{\G_{1s} n_1 (\beta_{rs, s}+\beta_{1s})}{\G_{2s} n_2 (\beta_{sf, s}+\beta_{2s})},
\eeq
where the speeds $\beta$'s are all defined positive and are measured in the shocked fluid rest frame; the subscript `$rs$' and `$fs$' refer to the RS and FS front, respectively. From the shock jump conditions (e.g., Blandford \& McKee 1976), one gets
\beq
\beta_{rs,s}= \frac{(\hat{\g}-1)\beta_{1s}}{1+1/\G_{1s}},
\eeq
 where $\hat{\g}$ is the adiabatic index for a relativistic fluid. Using an empirical relation $\hat{\g}= (4\bar{\G}+1)/(3\bar{\G})$ to smoothly connect the sub-relativistic shock regime to the relativistic shock regime, we obtain $(\beta_{rs, s}+\beta_{1s})= 4 \beta_{1s}/3$. Similar result applies to the FS front. Thus, Eq. (\ref{eq:Rm}) becomes
\beq    \label{eq:RF-Rm}
\frac{F_{\nu,\max,1}}{F_{\nu,\max,2}}= \frac{m_1}{m_2}= \frac{\G_{1s} \beta_{1s} n_1}{\G_{2s} \beta_{2s} n_2}= \left(\frac{n_1}{n_2}\right)^{1/2},
\eeq
where we have used Eq. (\ref{eq:p-bal}). This result is also numerically plotted in Figure \ref{fig:rsfs}.

%%%%%%%%%%%%%%%%%%%%%%%%%%%%%%%%%%%%%%%%%%%%%%%%%%%%%%%%%%%
%%%%%%%%%%%%%%%%%%%%%%%%%%%%%%%%%%%%%%%%%%%%%%%%%%%%%%%%%%%

According to Figure \ref{fig:rsfs}, in the internal RS-FS model, the optical emission shell must have a higher pre-shock density and a larger shock swept-up mass, hence should have a higher $F_{\nu,\max}$. The analysis of GRB 110205A prompt X/$\g$-ray spectrum suggests that the characteristic synchrotron frequencies are $h \nu_{c,\g} \sim$ 5 keV and $h \nu_{m,\g} \sim 300$ keV (see \S \ref{sec:analysis_prompt}); then we have $F_{\nu, \max, \g}/F_{\nu, opt} \sim 20$, and $\nu_{m,\g}/\nu_{opt} \sim 6\times10^4$. Therefore, if the internal RS-FS model would work for this burst, the maximum flux density of the optical producing shell $F_{\nu,\max}$ must be $\gg F_{\nu,opt}$; since the cooling frequencies of the two shells are equal (Eq. \ref{eq:Rnuc}), the optical shell must be in the slow cooling ($\nu_m < \nu_c$) regime.

In order to have the observed $F_{\nu,opt}$ much smaller than the optical shell $F_{\nu,\max}$, either $\nu_{opt}$ has to be far below or far above $\nu_m$, or the self-absorption frequency has to be $\nu_a > \nu_{opt}$, or both. In the following, we use the observed $F_{\nu, \max, \g}/F_{\nu, opt}$ and $\nu_{m,\g}/\nu_{opt}$ as constraints and $\nu_a/\nu_{opt}$ (for the optical shell) as a free parameter, and derive the permitted relation between $\nu_{m,\g}/\nu_{m,opt}$ and $n_{\g}/n_{opt}$, where $n_{\g}/n_{opt}$ is the density ratio of the $\g$-ray shell over the optical shell and is given by the maximum flux density ratio of the two shells (Eq. \ref{eq:RF-Rm}). We then overlay the permitted relations onto the internal RS-FS model predictions shown in Fig. \ref{fig:rsfs} right panel, in order to find the permitted model parameter values, i.e., shell LF ratio $\G_{fast}/\G_{slow}$, shell density ratio $n_{\g}/n_{opt}$ and $\nu_a/\nu_{opt}$.

\begin{figure}
\centerline{
\includegraphics[width=.59\textwidth, angle=0]{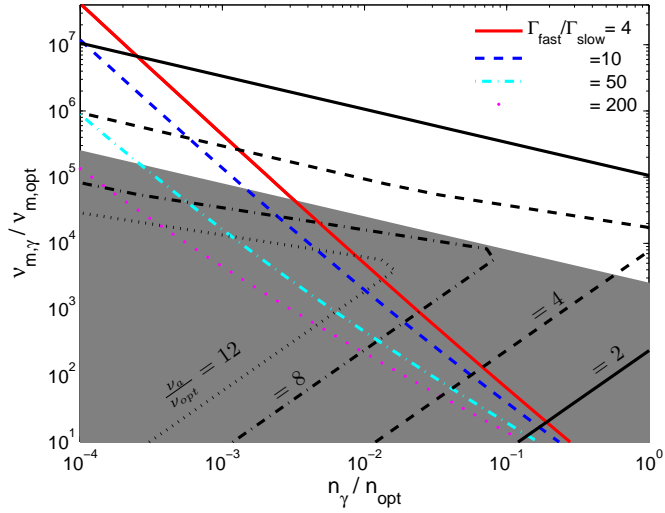}
}
\caption{The GRB 110205A-permitted $\nu_{m,\g}/\nu_{m,opt}$ versus $n_{\g}/n_{opt}$ relations for varying $\nu_a/\nu_{opt}$ values (\textit{black} lines), superimposed on the internal RS-FS model predictions (\textit{colored} lines, from the right panel of Fig. \ref{fig:rsfs}). The locations where the data-permitted and the model predicted relations intersect correspond to the specific model parameter values with which the model could work for GRB 110205A. Electron index $p= 3$ is assumed. The shaded region is forbidden because there the flux density at $\nu_X= 0.3$ keV in the emission spectrum of the optical producing shell will exceed the observed value of $F_{\nu,X}$, causing the spectral slope inconsistent with the observed one. Note that for $\nu_a < \nu_{opt}$ the data-permitted $\nu_{m,\g}/\nu_{m,opt}$ versus $n_{\g}/n_{opt}$ relations have no intersection with the internal RS-FS predictions unless $n_{\g}/n_{opt}$ is unreasonably small.}	 \label{fig:rsfs-nua}
\end{figure}

For the optical shell, the relation between $F_{\nu,opt}$ and $F_{\nu,\max}$ is determined according to the standard broken power law synchrotron spectrum (e.g. Sari et al. 1998; Granot \& Sari 2002), depending on the free parameter $\nu_a/\nu_{opt}$ which varies from $<1$ to $>1$. In addition, we impose an additional constraint that the high energy spectrum of the optical producing shell emission should not exceed the observed flux density at the lower bound of the X-ray band, i.e., at $\nu_X= 0.3$ keV, so that the spectral slope there would not be inconsistent with the observed one. The final results are shown in Figure \ref{fig:rsfs-nua}.

%%%%%%%%%%%%%%%%%%%%%%%%%%%%%%%%%%%%%%%%%%%%%%%%%%%%%%%%%%%
%%%%%%%%%%%%%%%%%%%%%%%%%%%%%%%%%%%%%%%%%%%%%%%%%%%%%%%%%%%

From Figure \ref{fig:rsfs-nua}, we conclude that the prompt SED data of GRB 110205A can be reproduced by the internal RS-FS synchrotron model under the following conditions: 
$\nu_{m,\g}/\nu_{m,opt} \approx 10^6 - 10^7$, 
$n_{\g}/n_{opt} \approx 10^{-4} - 10^{-3}$, and $\nu_a/\nu_{opt} \approx 2 - 6$ for the optical shell; 
and the LF ratio between the two shells falls into a wide range $\sim 4 - 100$. In Figure \ref{fig:rsfs-nua}, the electron index $p= 3$ has been adopted. A smaller $p$ value only increases the inferred $\nu_{m,\g}/\nu_{m,opt}$ and decreases $n_{\g}/n_{opt}$ both by a factor $< 10$, without affecting the conclusion.

\subsubsection{Emission radius}

The distance of the GRB emission region from the central engine ($R_{\rm GRB}$) has been poorly constrained. If prompt optical data are observed, one may apply the constraint on the synchrotron self-absorption frequency ($\nu_a$) to constrain $R_{\rm GRB}$ (Shen \& Zhang 2009). One needs to assume that the optical and $\g$-ray emission are from essentially the same radius in order to pose such a constraint. This is the simplest scenario, and is valid for some scenarios we have discussed, e.g. the internal RS-FS model as discussed in section \ref{sec:int_RS_FS}. In the following, we derive the emission radius based on the one-zone assumption, bearing in mind that optical and gamma-ray emissions can come from different zones.

For GRB 110205A, the synchrotron optical emission from the optically producing shell is self-absorbed and has the following frequency ordering: $\nu_m < \nu_{opt} < \nu_a < \nu_c$ (\S\ref{sec:int_RS_FS}).
For such a frequency ordering, $\nu_a$ is determined by (Shen \& Zhang 2009)
\beq
2 \g_a m_e \nu_a^2 = F_{\nu_a} \left(\frac{D_L}{R}\right)^2 \frac{(1+z)^3}{\G},
\eeq
where $D_L$ is the luminosity distance, $F_{\nu_a}$ is the observed flux density at $\nu_a$, and $\g_a= [16 m_e c (1+z) \nu_a/(3 e B' \G_s)]^{1/2}$ is the LF of electrons whose synchrotron frequency is $\nu_a$. Expressing $F_{\nu_a}$ in terms of the self-absorbed optical flux density: $F_{\nu_a}= F_{\nu,opt} (\nu_a/\nu_{opt})^{5/2}$, we find $\nu_a$ is canceled out in the above equation:
\beq
2 m_e \left(\frac{16 m_e c}{3 e B'}\right)^{1/2} \left(\frac{R}{D_L}\right)^2 \frac{(1+z)^{7/2}}{\G_s^{3/2}} = F_{\nu,opt} \nu_{opt}^{-5/2}.
\eeq
For an observed average $F_{\nu,opt} \approx 10^{1.8} \mu$Jy, $h\nu_{opt} \approx 10^{-2.3}$ keV and other numbers for GRB 110205A, we obtain
\beq
R_{\rm GRB} \approx 2.8\times10^{13}\, \left(\frac{\G_s}{250}\right)^{3/4} \left(\frac{B'}{10^5\, {\rm G}}\right)^{1/4}\, {\rm cm},
\eeq
We have normalized $\G_s$ to 250, and $B'$ to $10^5$ G. The former normalization can be justified from the afterglow data (\S\ref{sec:LF}).
The value of $B'$ is loosely determined, and may be estimated
$B' \sim 10^6~ \zeta L_{52} R_{13} \Gamma_2$ G, where
$\zeta \leq 1$ is a constant parameter (Zhang \& M\'esz\'aros 2002).
This gives $B' \sim 10^5$ G for the parameters of GRB 110205A.
Interpreting $E_1$ as $\nu_m$ would also give rise to  $B' \sim 10^5$ G
if $\Gamma_{1s} \sim 3$ and $\eps_e \sim 0.1$. We note that
$R_{\rm GRB}$ is a weak function of $B'$, so that an estimate $R_{\rm GRB} \approx 3\times10^{13}$ cm is
robust. This radius is consistent with the expectation of the
internal shock model (e.g. Piran 2005).

Note that in the Shen \& Zhang (2009) method one has to assume both the
optical and $\gamma$-rays belong to the same continuum component and
partly rely on the $\gamma$-ray low-energy spectral slope to constrain the
$\nu_a$ location. However, this assumption is discarded in the case of GRB
110205A: it is inferred that in the internal RS-FS model the optical shell
has $\nu_{opt} < \nu_a$ (\S \ref{sec:int_RS_FS}), for which $R_{GRB}$ is worked out
without the aid of the $\g$-ray spectral information.

\subsubsection{Two-zone models}

The GRB 110205A prompt observations might be also interpreted if the optical emission region is decoupled from the $\g$-ray/X-ray emission regions. There are models which envisage that the $\g$-rays are produced in internal shocks at small radii between shells with large LF contrasts, while the optical emission is generated in internal shocks at larger radii by shells with lower magnetic fields and smaller LF contrasts. This can happen in two scenarios. According to Li \& Waxman (2008), even after collisions and mergers of shells with large LF contrasts, the outflow still comprises discrete shells with variations, though with reduced relative LFs, which could lead to the ``residual'' collisions at larger radii. Fan et al. (2009) considered a neutron-rich GRB outflow, in which free neutrons are decoupled from the proton shells until decaying at large radii. Violent collisions among the proton shells occur at smaller radii, while some later-ejected, slower, proton shells catch up with the decayed neutron shells at large radii and give weaker collisions. Both scenarios might work for GRB 110205A, since in either case, the large-radii collisions would bear a similar temporal information as the small-radii collisions, rendering a coarse optical-$\g$-ray correlation.
A defining property of these two-zone scenarios is that the optical pulses should display a larger variability time scale $\delta t$ than the $\g$-ray pulses, and they should lag behind the $\g$-ray pulses by $\sim R_{opt}/2\Gamma^2 c \sim 0.2~{\rm s} R_{opt,15} \Gamma_{2.5}^{-2}$ s. These predictions could be tested by future, high temporal-resolution, prompt optical observations of similar bursts.

\subsubsection{Dissipative photosphere emission model}

Recently, several independent groups (e.g., Giannios 2008; Toma et al. 2010; Beloborodov 2010; Lazzati \& Begelman 2010; Vurm et al. 2011) have developed an improved version of the photosphere emission model of GRB prompt emission. This model invokes energy dissipation in the Thomson-thin layer of the photosphere, so that the photosphere spectrum deviates from the thermal form through IC upscattering. The same electrons also emit synchrotron photons, which may account for the optical excess. A difficulty of this scenario is that the low energy spectral index below $E_p$ is too hard (e.g. $\alpha=+0.4$, Beloborodov 2010) to account for the observations ($\alpha=-1.5$). The data of GRB 110205A (double breaks and spectral slopes) of the X/$\g$-ray component do not comply with the predictions of this model, but is rather consistent with the the standard synchrotron model 
(see section \ref{sec:general-prompt}).
We conclude that the dissipative photosphere model does not apply at least to this burst.

\subsection{Afterglow modeling\label{sec:modeling_afterglow}}

\subsubsection{Initial Lorentz factor}
\label{sec:LF}

For both scenarios I and II, the optical peak time $t_p$ = 1045 $\pm$ 63 s
corresponds to the deceleration time. Since this time is much longer than
$T_{90}$ ($\sim 257$ s), it is pertinent to consider a ``thin'' shell regime
(Sari \& Piran 1995). The peak time can be then used to estimate the
initial Lorentz factor of the ejecta
(e.g. Meszaros 2006; Molinari et al. 2007):
$\Gamma_0 \sim 560
({E_{\gamma,iso,52}}/{\eta_{0.2}n_0t^3_{p,z,1}})^{1/8}$,
where E$_{\gamma,iso,52}$ is the isotropic equivalent energy in units of
10$^{52}$ erg s$^{-1}$; $\eta_{0.2}$ is the radiative efficiency in
units of 0.2; $n_0$ is local density in units of cm$^{-3}$ and
$t_{p,z,1}$ is the peak time corrected for cosmological time
dilation in units of 10 s. For GRB 110205A, with redshift $z$ = 2.22
and fluence of 2.7$^{+0.7}_{-0.4} \times$ 10$^{-5}$ erg cm$^{-2}$
(15 - 3000 keV, Sakamoto et al. 2011c), we derive the rest-frame
1 keV - 10 MeV isotropic energy  E$_{\gamma,iso}$ = 46$^{+4}_{-7}$
$\times$ 10$^{52}$ erg. With $t_{p,z,1}$ = 32.4$\pm$1.8, we finally
estimate
\begin{equation}
\Gamma_0 = 245^{+7}_{-10}(\eta_{0.2}n_0)^{-1/8}.
\end{equation}
This value follows the
empirical relation $\Gamma_0$ $\sim$ 182E$^{0.25\pm0.03}_{\gamma,iso,52}$
recently found by Liang et al. (2010) within 2-$\sigma$ range.

\subsubsection{Light curves}

Scenario I :

In this scenario, the optical bump around $t_p$ is mostly contributed by
the FS. Fitting the light curves, one can constrain the temporal slopes
of the RS component. In the $R$-band, the temporal indices are
$\alpha_{r1}$ =3.32$\pm$1.2 (rising phase) and $\alpha_{r2}$ = -5.90$\pm$1.0
(decaying phase), while in the X-ray band, they are
$\alpha_{r1}$ = 5.19$\pm$1.3 (rising) and $\alpha_{r2}$ =
-8.26$\pm$1.3 (decaying).
The steep rising slope ($\sim 4$) is consistent with the expectation in the
thin shell ISM RS model (e.g. Kobayashi 2000, Zhang et al. 2003).
The decaying slopes look too steep as compared with the theoretically
expected values (e.g. $\sim -2-\beta$ for the so-called
``curvature'' effect, Kumar \& Panaitescu 2000). However, strictly
speaking, the expected decay index is valid when the time zero point
is placed to $t_p$. The results are therefore not inconsistent with the
theoretical expectations.  Compared with other GRBs with the RS emission
identified (which typically peaks around or shortly
after the $T_{90}$ duration), a RS emission peaking at $\sim$ 1100 s
after the burst is rare and has not been seen before (though previously, the optical brightening in the ultra-long GRB 091024 was claimed by Gruber et al. 2011 to be caused by the rising RS, but its data coverage is very sparse
and the RS origin was not exclusively determined, e.g., it could also be due to the FS peak in a wind medium).

For the FS component, the rising index is set to +3 during the
fitting for both the optical and X-ray bands. The decay index after
the peak is fitted to $\alpha_{of2}$ =
-1.50$\pm$0.04 in the optical band, and
$\alpha_{Xf2}$ = -1.54$\pm$0.1 in the X-ray band.
We also constructed two SEDs at the epochs 5.9 ks and 35 ks during the
decay phase.
We find that the only model that satisfies the closure relation (e.g.
Sari et al. 1998; Granot \& Sari 2002; Zhang \& M\'esz\'aros 2004)
is the ISM model in the $\nu_m < \nu_o
< \nu_x < \nu_c$ spectral regime. For example, our SED at 5.9 ks
gives a spectral slope $\beta \sim -1.01 \pm 0.02$ across the entire
energy band. The optical temporal decay index $\alpha_o =
-1.50\pm 0.04$ matches the expected closure relation $\alpha =
(3/2)\beta$ well.
The X-ray decay slope $\alpha_{Xf2}$ = -1.54$\pm$0.1
within 1-$\sigma$ error is consistent with the closure relation.
The electron energy index, $p = -2 \beta+1 \sim 3.0$, is also
consistent with its value derived from the temporal index
index $p = (-4\alpha+3)/3  = 3.0 \pm
0.08$ derived from $\alpha_o$ and $p = 3.05 \pm 0.14$ derived from
$\alpha_x$.

At late times around $\sim 10^5$ s, the decay index becomes
steeper in both the optical and X-ray bands, which is probably caused
by a jet break. The simultaneous fit suggests a $t_{jb} = 1.0 \pm 0.2 \times 10^5$ s
from the two band light curves. The post-break temporal indices are consistent
with a jet break without significant sideways expansion, which is
predicted to be steeper by 3/4. But the conclusion is not
conclusive due to large errors. 
The jet angle can be calculated using (Sari et al. 1999)
$\theta$ = $\frac{1}{6}
\big{(}\frac{t_{jb}}{1+z}\big{)}^{3/8}
\big{(}\frac{n\eta_{0.2}}{E_{\gamma,iso,52}}\big{)}^{1/8}$.
Taking $t_{jb} \simeq 1.0 \times 10^5$ s = 1.2 days,
and $E_{\gamma,iso,52}$ = 46.0, we derive $\theta$ = (4.1$^{+3.5}_{-1.0}$)$^\circ$.
With a beaming factor of $\frac{\theta^2}{2}$, the
corresponding jet-angle-corrected energy is
$E_{\gamma}$ = 1.2 $\times$ 10$^{51}$ erg.

Scenario II :

In this scenario, the early steep rise and bright peak is
dominated by the RS component only. It can also be well explained by
a ISM model in the thin shell regime. Within this scenario, the FS
component shows up and peaks later. The FS peak is defined by
$\nu_m$ crossing the optical band. There should also be a break
time in the FS light curve at the RS peak $t_p$, which is caused by the
onset of the afterglow. After the FS peak, the afterglow analysis
is similar to Scenario I. We
find the best afterglow model for this scenario is the ISM model
assuming $\nu_m <  \nu_c <  \nu_o <  \nu_x$.
From the spectral index $\beta$ = -1.01, one can derive $p= 2.02$.
Since $\alpha \sim$ 1.0 for both optical and X-ray,  the closure
relation $\alpha$ = (3$\beta$+1)/2 is well satisfied.

The X-ray bump around 1000 s is also consistent with being the the emission from the RS.
Our SED fit (see Table \ref{tab:sed_ref}) near the peak at 1.13 ks shows that the best
fits are a single power-law (LMC and SMC model), or a broken power-law model (LMC model)
with the break energy $\sim$ 2.9 keV. The result suggests that the optical and
X-ray bands belong to a same emission component with $\nu_m < \nu_{opt} < \nu_X < \nu_c$
(for single power law) or $\nu_m < \nu_{opt} < \nu_c < \nu_X$ (for broken power law).
In the case of a thin ejecta shell and ISM,
the synchrotron cooling frequency at the time when the RS
crosses the ejecta, $t_p$, can be estimated to be 
$\nu_c (t_p)= 2 \times 10^{26} \Gamma^{-4}$ ($\epsilon_B n)^{-3/2}
(1+z) t_p^{-2}$ Hz (also see Kobayashi 2000). Adopting $\Gamma \sim 250$, $z= 2.22$, 
$t_p= 1.1$ ks that are relevant for GRB 110205A, this would require 
$\epsilon_B n \sim 10^{-5}$ cm$^{-3}$.
Such a value, although in the low end of the generally anticipated parameter distribution
range, is not impossible.

The jet break time derived from this scenario is somewhat earlier than
Scenario I, which is $\sim 5.44 \pm 0.2 \times 10^4$ s from the
simutaneous fitting. Adopting this break time we derived the jet angle for
this scenario is $\theta =
({3.2^{+2.6}_{-0.9}}$)$^\circ$, which corresponds to a jet-angle-corrected
energy of $E_{\gamma} = 7.2 \times 10^{50}$ erg.

\subsubsection{RS magnetization}

The composition of the GRB ejecta is still not well constrained.
Zhang et al. (2003) suggested that bright optical flashes generally
require that the RS is more magnetized than the FS, namely, the
ejecta should carry a magnetic field flux along with the matter
flux (see also Fan et al. 2002, Panaitescu \& Kumar 2004 for case
studies). Since GRB 110205A has a bright RS component, it is interesting
to constrain the RS parameters to see whether it is also magnetized.
For both scenarios I and II, since the peak time and maximum flux of
both FS and RS can be determined (also shown in Table
 \ref{tab:fit_res_R_X_lc}), one can work out the constraints on the
RS magnetization following the method delineated in Zhang et al.
(2003). The same notations are adopted here as in Zhang et al. (2003).

For Scenario I, the FS peaks at $t_{\times}$ where $\nu_{m,f} < \nu_R < \nu_{c,f}$, and then decays as $F_{\nu,f} \propto t^{-3(p_f-1)/4}$.
For Scenario II, In the FS, one has $\nu_R \leq \nu_{m,f} < \nu_{c,f}$ at $t_{\times}$. The FS light curve still rises as $t^{1/2}$, until reaching $t_{p,f}$ where $\nu_R= \nu_{m,f} < \nu_{c,f}$.
We define ($t_{p,r}$, $F_{\nu, p, r}$) and ($t_{p,f}$, $F_{\nu, p, f}$) as the peak times and peak flux densities in optical for RS and FS, respectively. Similar to those presented in Zhang et al. (2003), it follows naturally from the above that
\begin{equation}    \label{eq:Re_t}
\Re_t \equiv \frac{t_{p,f}}{t_{p,r}}= \begin{cases}
1, & {\rm for~ scenario~ I}\\
\G_{\times}^{4/3} \Re_B^{-2/3} \Re_{\nu}^{-2/3}, & {\rm for~ scenario~ II}
\end{cases}
\end{equation}
\begin{equation}    \label{eq:Re_F}
\Re_F \equiv \frac{F_{\nu, p ,r}}{F_{\nu, p, f}}= \begin{cases}
\G_{\times}^{2-p_f} \Re_B^{\frac{p_f+1}{2}} \Re_{\nu}^{\frac{p_f-p_r}{2}}, & {\rm for~ scenario~ I}\\
\G_{\times} \Re_B \Re_{\nu}^{-\frac{p_r-1}{2}}, & {\rm for~ scenario~ II}
\end{cases}
\end{equation}
Note that Scenario I is actually a special case of $\Re_t= 1$ of the more general Scenario II. The results are identical to those in Zhang et al. (2003) except that we do not use the RS decay slope $\alpha_{r,2}$ after $t_{\times}$ as one of the parameters.
This was to avoid the ambiguity of the blastwave dynamics after the shock crossing. Instead we keep $p_r$ in the formulae, and determine $p_r$ from the better understood RS rising slope before $t_{\times}$: $\alpha_{r,1}= (6p_r-3)/2$.

For Scenario I, one can derive $\Re_t \sim 1$, $p_r \approx 2$, $p_f \approx 3$ and $\Re_F \approx 0.5$. Notice that the entire optical peak around $t_{\times}$ is contributed mainly from the FS, which means $p_r$ is poorly constrained. So we take $p_r = p_f \approx 3$. Then from Eq. (\ref{eq:Re_F}) we get
\begin{equation}    \label{eq:Re_B1}
\Re_B \sim 7~ \G_{\times,2}^{1/2} \sim 7.7.
\end{equation}

For Scenario II, $\Re_t \approx 3.6$, $\Re_F \approx 16.5$ and $p_r \approx 2.3$. Plugging in numbers and from Eq. (\ref{eq:Re_t}) we have $\Re_B \Re_{\nu} \sim \G_{\times}^2/6.8$. Combining it with Eq. (\ref{eq:Re_F}) and canceling out $\Re_{\nu}$, we get
\begin{equation}    \label{eq:Re_B2}
\Re_B \sim 7~ \G_{\times,2}^{0.18} \sim 7.2.
\end{equation}
The numerical values are obtained by taking $\Gamma_{\times} = \Gamma_0/2 \sim 120$.

In both scenarios, the magnetic field strength ratio $\Re_B \equiv B_r/B_f$ is $\sim 7$.
This suggests that the RS is more magnetized than the FS.
Since the magnetic field in the FS is believed to be induced by plasma
instabilities (Weibel 1959; Medvedev \& Loeb 1999; Nishikawa et al. 2009),
a stronger magnetic field in the RS region must have a primordial origin,
i.e. from a magnetized central engine.

\subsubsection{Discussion}
Comparing with the other recent work on GRB 110205A (Cucchiara et al. 2011,
Gao 2011 and Gendre et al. 2011), our scenario II analysis, which concludes
the bright optical peak around 1000 s is dominated by the RS emission, agrees with
that by Gendre et al. (2011) and Gao (2011). Our scenario I
analysis, though close to the conlsuison by Cucchiara that it is dominated
the FS emission, has slight difference, as we also consider the RS contribution
in our scenario I.

Both scenarios in our analysis can interpret the general properties of
the broadband afterglow. However, each scenario has some caveats.
For Scenario I, as explained above, the best fit model light curve
is not steep enough to account for the data in the UVOT-$u$ band
( $\alpha \sim$ 5 for $R$ and $\alpha \sim$ 6 for
UVOT-$u$ if apply a single broken-power-law fitting).
Since the inconsistency only occurs in the bluest band with adequate
data to constrain the rising slope, we speculate that the steeper
rising slope may be caused by a decreasing extinction with time
near the GRB. No clear evidence of the changing extinction has been
observed in other GRBs. In any case, theoretical models have
suggested that dust can be destructed by strong GRB X-ray and UV
flashes along the line of sight, so that a time-variable extinction
is not impossible (e.g. Waxman \& Draine 2000;
Fruchter et al. 2001; Lazzati et al. 2002a; De Pasquale et al. 2003).
For Scenario II, the model cannot well fit the X-ray peak around
1100 s. The main reason is that the required RS component to fit
the optical light curve is not as narrow as that invoked in
Scenario I. It is possible that X-ray feature is simply an X-ray
flare due to late central engine activities, which have been
observed in many GRBs (e.g. Burrows et al. 2005b; Liang et al. 2006;
Chincarini et al. 2007).

In the late optical light curve around 5$\times$10$^4$ s, there is a
re-brightening bump observed by LOT in four bands. Such bumps have
been seen in many GRBs (e.g. GRB 970508, Galama et al. 1998; GRB 021004, Lazzati et al. 2002b; GRB
050820, Cenko et al. 2006; GRB 071025, Updike et al. 2008; GRB 100219A, Mao et al. 2011),
which are likely caused by the medium density bumps (e.g.
Lazzati et al. 2002b;
Dai \& Wu 2003; Nakar \& Granot 2007; Kong et al. 2010).
Microlensing is another possibility (e.g. Garnavich et al. 2000;
Gaudi et al. 2001; Baltz \& Hui 2005), although the event rate is
expected to be rather low.

Interestingly, linear polarization at a level of P $\sim$ 1.4\%
was measured by the 2.2 m telescope at Calar Alto Observatory
(Gorosabel et al. 2011) 2.73-4.33 hours after the burst. During this
time, the afterglow is totally dominated by the FS in Scenario
I, or mostly dominated by the FS (with a small contamination from
the RS) in Scenario II.
The measured linear polarization degree is similar to several
other detections in the late afterglow phase (e.g. Covino et al.
1999, 2003; Greiner et al. 2003; Efimov et al. 2003), which is consistent with the
theoretical expectation of synchrotron emission in external shocks
(e.g. Gruzinov \& Waxman 1999; Sari 1999; Ghisellini \& Lazzati 1999).

\section{Summary}

We have presented a detailed analysis of the bright GRB 110205A, which was detected by both $Swift$ and $Suzaku$. Thanks to its long duration, $Swift$ XRT, UVOT, ROTSE-IIIb and BOOTES telescopes were able to observe when the burst was still in the prompt emission phase. Broad-band simultaneous observations are available for nearly 200 s, which makes it one of the exceptional opportunities to study the spectral energy distribution during the prompt phase. The broad-band time-resolved spectra are well studied. For the first time, an interesting two-break energy spectrum is identified throughout the observed energy range, which is roughly consistent with the synchrotron emission spectrum predicted by the standard GRB internal shock model.
Shortly after the prompt emission phase, the optical light curve shows a bump feature around 1100 s with an unusual steep rise ($\alpha$ $\sim$ 5.5) and a bright peak ($R$ $\sim$ 14.0 mag). The X-ray band shows a bump feature around the same time. This is followed by a more normal decay behavior in both optical and X-ray bands. At late times, a further steepening break is visible in both bands.

The rich data in both the prompt emission and afterglow phase make GRB 110205A an ideal burst to study GRB physics, to allow the study of the emission mechanisms of GRB prompt emission and afterglow, and to constrain a set of parameters that are usually difficult to derive from the data. It turns out that the burst can be well interpreted within the standard fireball shock model, making it a ``textbook'' GRB. We summarize our conclusions as follows.

1. The two-break energy spectrum is highly consistent with the synchrotron emission model in the fast cooling regime. This is consistent with the internal shock model or the magnetic dissipation model that invokes first-order Fermi acceleration of electrons.

2. The prompt optical emission is $\sim 20$ times greater than the extrapolation from the X/$\g$-ray spectrum. Our analysis rules out the synchrotron + SSC model to interpret the optical + X/$\g$-ray emission.
We find that the prompt emission can be explained by a pair of reverse/forward shocks naturally arising from the conventional internal shock model. In a two-shell collision, the synchrotron emission from the slower shock that enters the denser shell produces optical emission and is self-absorbed while that from the faster shock entering the less dense shell produces the X/$\g$-ray emission. The required density ratio of two shells is $\sim 10^{-4} - 10^{-3}$.

3. If the optical and gamma-ray emissions originate from a same radius, as is expected in the internal forward/reverse shock model, one can pinpoint the prompt emission radius to $R \sim 3\times 10^{13}$ cm by requiring that the synchrotron optical photons are self-absorbed.

4. The data can be also interpreted within a two-zone model where X/$\g$-rays are from a near zone, while the optical emission is from a far zone. The dissipative photosphere model is inconsistent with the prompt emission data.

5. The broad band afterglow can be interpreted within the standard RS + FS model. Two scenarios are possible: Scenario I invokes both FS and RS to peak at 1100 s, while Scenario II invokes RS only to peak at 1100 s, with the FS peak later when $\nu_m$ cross the optical band. In any case, this is the first time when a rising reverse shock -- \textit{before} its passage of the GRB ejecta (not after, when the reverse shock emission is fast decaying, like in GRB 990123 and a few other cases) -- was observed in great detail.

6. In either scenario, the optical peak time can be used to estimate the initial Lorentz factor of GRB ejecta, which is found to be $\G_0 \approx 250$.

7. From the RS/FS modeling, we infer that the magnetic field strength ratio in reverse and forward shocks is $\Re_B \equiv B_r/B_f$ $\sim 7$. This suggests that the GRB ejecta carries a magnetic flux from the central engine.

8. Jet break modeling reveals that the GRB ejecta is collimated, with an opening angle $\sim 4.1^{\circ}$ (Scenario I) or $\sim 3.2^{\circ}$ (Scenario II). The jet-corrected $\g$-ray energy is $E_\gamma \sim 1.2 \times 10^{51}$ erg or $E_\gamma \sim 7.2\times 10^{50}$ erg.

\vspace{0.1cm}

\acknowledgments
We thank the anonymous referee for helpful comments and suggestions to improve the manuscript.
This research is supported by the NASA grant NNX08AV63G and the NSF
grant PHY-0801007. RFS is supported by an NSERC Discovery grant.
APB, AAB, NPK, MJP and SRO acknowledges the support from the UK Space Agency.
BZ acknowledges NASA NNX10AD48G and NSF AST-0908362 for support.
MI acknowldeges support from the CRI grant 2009-0063616, funded by MEST of the Korean government.
The Centre for All-sky Astrophysics is an Australian Research Council
Centre of Excellence, funded by grant CE11E0090. This research made use
of public data supplied by the High Energy Astrophysics Science Archive
Research Center (HEASARC) at the NASA Goddard Space Flight Center.
This work has been supported by Spanish Junta de Andaluc\'{\i}a through
program FQM-02192 and from the Spanish Ministry of Science and Innovation
through Proyects (including FEDER funds) AYA 2009-14000-C03-01 and
AYA2008-03467/ESP. We thank for support provided by INTA and EELM-CSIC for hosting
the BOOTES observatories. The work based partly on data acquired at the Centro
Astron\'omico Hispano Alem\'an (CAHA) de Calar Alto and Observatorio de
Sierra Nevada (OSN). This research was also supported by the UK STFC.

\begin{deluxetable}{rrrrccrrrrc}
 \tablewidth{0pt}
 \tablecaption{Photomertric observations for GRB 110205A from ground-based telescopes\label{tab:Opt_photometry_data}}
  \tablehead{\colhead{T-T$_0$ (s)} & \colhead{Exp (s)} & \colhead{Mag} & \colhead{Error} & \colhead{Filter} & \colhead{} & \colhead{T-T$_0$ (s)} & \colhead{Exp (s)} & \colhead{Mag} & \colhead{Error} & \colhead{Filter}}
\startdata
%ROTSE-IIIb
\multicolumn{11}{c}{ROTSE-IIIb} \\
\hline
135.5            &  107.0   &  17.79  &  0.42  &  $C$  &  &  1695.7    &  60.0    &  14.84  &  0.06   &  $C$  \\
340.2            &  282.2   &  16.42  &  0.11  &  $C$  &  &  1765.0    &  60.0    &  14.89  &  0.08   &  $C$  \\
520.5            &  60.0    &  16.33  &  0.20  &  $C$  &  &  1834.2    &  60.0    &  15.04  &  0.09   &  $C$  \\
589.6            &  60.0    &  15.75  &  0.12  &  $C$  &  &  1903.0    &  60.0    &  15.01  &  0.07   &  $C$  \\
658.4            &  60.0    &  15.25  &  0.09  &  $C$  &  &  1971.8    &  60.0    &  15.22  &  0.08   &  $C$  \\
727.5            &  60.0    &  14.81  &  0.06  &  $C$  &  &  2040.7    &  60.0    &  15.28  &  0.08   &  $C$  \\
796.6            &  60.0    &  14.50  &  0.06  &  $C$  &  &  2109.5    &  60.0    &  15.23  &  0.07   &  $C$  \\
865.7            &  60.0    &  14.27  &  0.05  &  $C$  &  &  2178.4    &  60.0    &  15.44  &  0.11   &  $C$  \\
935.1            &  60.0    &  14.19  &  0.04  &  $C$  &  &  2247.2    &  60.0    &  15.40  &  0.11   &  $C$  \\
1004.3           &  60.0    &  14.08  &  0.03  &  $C$  &  &  2384.7    &  60.0    &  15.51  &  0.11   &  $C$  \\
1073.5           &  60.0    &  14.06  &  0.03  &  $C$  &  &  2453.9    &  60.0    &  15.46  &  0.11   &  $C$  \\
1142.7           &  60.0    &  14.14  &  0.05  &  $C$  &  &  2660.7    &  60.0    &  15.71  &  0.11   &  $C$  \\
1211.9           &  60.0    &  14.18  &  0.04  &  $C$  &  &  2868.0    &  60.0    &  15.83  &  0.12   &  $C$  \\
1281.0           &  60.0    &  14.31  &  0.06  &  $C$  &  &  2937.2    &  60.0    &  16.02  &  0.13   &  $C$  \\
1350.3           &  60.0    &  14.40  &  0.06  &  $C$  &  &  3075.3    &  60.0    &  15.82  &  0.15   &  $C$  \\
1419.5           &  60.0    &  14.53  &  0.05  &  $C$  &  &  3144.5    &  60.0    &  15.96  &  0.15   &  $C$  \\
1489.1           &  60.0    &  14.53  &  0.05  &  $C$  &  &  3213.4    &  60.0    &  16.00  &  0.13   &  $C$  \\
1558.0           &  60.0    &  14.55  &  0.06  &  $C$  &  &  3593.1    &  680.0   &  16.15  &  0.05   &  $C$  \\
1626.6           &  60.0    &  14.67  &  0.06  &  $C$  &  &  -         &  -       &  -      &  -      &  $-$  \\
%LOT
\\
\multicolumn{11}{c}{LOT (AB magnitude)} \\
\hline
39647.6          &  180.0   &  20.95  &  0.07  &  $g'$  &  &  43061.7   &  300.0   &  19.49  &  0.06   &  $z'$  \\
42062.9          &  300.0   &  21.05  &  0.05  &  $g'$  &  &  39856.7   &  180.0   &  20.32  &  0.05   &  $r'$  \\
45217.0          &  600.0   &  20.93  &  0.03  &  $g'$  &  &  42392.9   &  300.0   &  20.47  &  0.04   &  $r'$  \\
48824.2          &  600.0   &  20.90  &  0.02  &  $g'$  &  &  45846.0   &  600.0   &  20.35  &  0.04   &  $r'$  \\
52111.7          &  600.0   &  21.07  &  0.03  &  $g'$  &  &  49453.2   &  600.0   &  20.42  &  0.03   &  $r'$  \\
55478.8          &  600.0   &  21.25  &  0.03  &  $g'$  &  &  52740.7   &  600.0   &  20.51  &  0.02   &  $r'$  \\
58774.9          &  600.0   &  21.33  &  0.04  &  $g'$  &  &  56108.6   &  600.0   &  20.70  &  0.03   &  $r'$  \\
62047.7          &  600.0   &  21.44  &  0.04  &  $g'$  &  &  59404.8   &  600.0   &  20.80  &  0.03   &  $r'$  \\
65421.7          &  600.0   &  21.55  &  0.04  &  $g'$  &  &  62677.6   &  600.0   &  20.97  &  0.04   &  $r'$  \\
40064.1          &  180.0   &  19.97  &  0.06  &  $i'$  &  &  66052.4   &  600.0   &  20.98  &  0.03   &  $r'$  \\
42720.4          &  300.0   &  20.19  &  0.05  &  $i'$  &  &  68767.1   &  600.0   &  21.08  &  0.04   &  $r'$  \\
46474.1          &  600.0   &  20.11  &  0.03  &  $i'$  &  &  69377.1   &  600.0   &  21.10  &  0.04   &  $r'$  \\
50079.6          &  600.0   &  20.13  &  0.03  &  $i'$  &  &  69987.9   &  600.0   &  21.22  &  0.05   &  $r'$  \\
53368.0          &  600.0   &  20.31  &  0.03  &  $i'$  &  &  129844.0  &  600.0   &  22.50  &  0.17   &  $r'$  \\
56735.0          &  600.0   &  20.45  &  0.04  &  $i'$  &  &  130563.0  &  600.0   &  22.54  &  0.15   &  $r'$  \\
60032.0          &  600.0   &  20.62  &  0.04  &  $i'$  &  &  131174.0  &  600.0   &  22.36  &  0.11   &  $r'$  \\
63304.9          &  600.0   &  20.69  &  0.05  &  $i'$  &  &  133798.0  &  600.0   &  22.47  &  0.09   &  $r'$  \\
66678.8          &  600.0   &  20.81  &  0.04  &  $i'$  &  &  134409.0  &  600.0   &  22.38  &  0.09   &  $r'$  \\
40287.9          &  180.0   &  19.68  &  0.09  &  $z'$  &  &  135019.0  &  600.0   &  22.60  &  0.12   &  $r'$  \\
%HCT
\\
\multicolumn{11}{c}{HCT} \\
\hline
67639.0          &  1080.0  &  21.00  &  0.04  &  $R $  &  &  72379.0   &  600.0   &  21.15  &  0.04   &  $R $  \\
69019.0          &  540.0   &  21.00  &  0.05  &  $R $  &  &  -         &  -       &  -      &  -      &  $- $  \\
%Lightbuckets
\\
\multicolumn{11}{c}{0.61-m Lightbuckets} \\
\hline
20747.0          &  300.0   &  18.67  &  0.02  &  $C$  &  &  27549.9   &  300.0   &  19.19  &  0.08   &  $R$  \\
%GRAS
\\
\multicolumn{11}{c}{GRAS 005} \\
\hline
3536.8           &  600.0   &  15.98  &  0.02  &  $C$  &  &  4248.7    &  600.0   &  16.44  &  0.02   &  $R$  \\
5327.4           &  300.0   &  16.51  &  0.02  &  $C$  &  &  5857.5    &  600.0   &  16.89  &  0.03   &  $R$  \\
6902.9           &  600.0   &  17.02  &  0.03  &  $C$  &  &  6378.8    &  300.0   &  17.10  &  0.04   &  $R$  \\
8366.9           &  300.0   &  17.33  &  0.04  &  $C$  &  &  7992.8    &  300.0   &  17.62  &  0.08   &  $R$  \\
%OSN
\\
\multicolumn{11}{c}{1.5-m OSN} \\
\hline
10152.2          &  300.0   &  18.93  &  0.03  &  $B$  &  &  12747.2   &  90.0    &  18.15  &  0.02   &  $R$  \\
11527.2          &  300.0   &  19.07  &  0.03  &  $B$  &  &  13337.9   &  90.0    &  18.18  &  0.03   &  $R$  \\
12426.7          &  150.0   &  19.17  &  0.03  &  $B$  &  &  14361.8   &  90.0    &  18.34  &  0.03   &  $R$  \\
13091.5          &  150.0   &  19.21  &  0.03  &  $B$  &  &  14856.1   &  90.0    &  18.32  &  0.03   &  $R$  \\
14115.7          &  150.0   &  19.32  &  0.03  &  $B$  &  &  85598.9   &  3300.0  &  20.89  &  0.05   &  $R$  \\
14610.0          &  150.0   &  19.38  &  0.04  &  $B$  &  &  10489.4   &  300.0   &  18.67  &  0.06   &  $U$  \\
76574.1          &  400.0   &  21.87  &  0.17  &  $B$  &  &  11871.4   &  500.0   &  18.96  &  0.06   &  $U$  \\
9871.3           &  240.0   &  17.22  &  0.02  &  $I$  &  &  13592.3   &  500.0   &  19.22  &  0.06   &  $U$  \\
11278.2          &  240.0   &  17.41  &  0.02  &  $I$  &  &  9644.8    &  180.0   &  18.04  &  0.02   &  $V$  \\
12840.2          &  200.0   &  17.59  &  0.02  &  $I$  &  &  11094.4   &  180.0   &  18.35  &  0.02   &  $V$  \\
13430.8          &  120.0   &  17.65  &  0.03  &  $I$  &  &  12653.7   &  90.0    &  18.55  &  0.02   &  $V$  \\
14454.7          &  120.0   &  17.77  &  0.03  &  $I$  &  &  13244.3   &  90.0    &  18.61  &  0.03   &  $V$  \\
14949.0          &  120.0   &  17.78  &  0.03  &  $I$  &  &  14268.5   &  90.0    &  18.71  &  0.04   &  $V$  \\
77630.9          &  300.0   &  20.56  &  0.19  &  $I$  &  &  14762.8   &  90.0    &  18.66  &  0.04   &  $V$  \\
9409.4           &  180.0   &  17.59  &  0.02  &  $R$  &  &  76977.2   &  350.0   &  21.68  &  0.19   &  $V$  \\
%80289.9   &  300.0   &  99.99  &  21.70  &  $U$  \\
%LOAO
\\
\multicolumn{11}{c}{1-m LOAO} \\
\hline
13101.0          &  720.0   &  19.32  &  0.16  &  $B$  &  &  12846.0   &  180.0   &  18.16  &  0.03   &  $R$  \\
11974.0          &  180.0   &  18.14  &  0.03  &  $R$  &  &  13273.0   &  180.0   &  18.22  &  0.03   &  $R$  \\
12416.0          &  180.0   &  18.11  &  0.03  &  $R$  &  &  -         &  -       &  -      &  -      &  $-$  \\
%BOOTES-1
\\
\multicolumn{11}{c}{BOOTES-1} \\
\hline
219.4            &  117.0   &  17.43  &  0.38  &  $C$  &  &  2715.2    &  49.0    &  15.66  &  0.07   &  $C$  \\
1603.5           &  48.0    &  14.61  &  0.05  &  $C$  &  &  2832.0    &  64.0    &  15.74  &  0.07   &  $C$  \\
2118.2           &  48.0    &  15.18  &  0.05  &  $C$  &  &  3096.5    &  197.0   &  15.81  &  0.04   &  $C$  \\
2218.1           &  48.5    &  15.29  &  0.05  &  $C$  &  &  3709.1    &  413.0   &  16.21  &  0.04   &  $C$  \\
2317.5           &  48.0    &  15.45  &  0.06  &  $C$  &  &  4621.4    &  495.5   &  16.54  &  0.05   &  $C$  \\
2417.5           &  48.5    &  15.44  &  0.06  &  $C$  &  &  5615.5    &  495.5   &  16.88  &  0.08   &  $C$  \\
2516.8           &  48.0    &  15.35  &  0.05  &  $C$  &  &  6658.3    &  545.0   &  17.15  &  0.08   &  $C$  \\
2615.8           &  48.0    &  15.51  &  0.07  &  $C$  &  &  8066.6    &  859.5   &  17.59  &  0.10   &  $C$  \\
%BOOTES-2
\\
\multicolumn{11}{c}{BOOTES-2} \\
\hline
939.3            &  5.0     &  14.40  &  0.04  &  $R$  &  &  6300.8    &  363.5   &  17.19  &  0.04   &  $R$  \\
994.7            &  5.0     &  14.24  &  0.04  &  $R$  &  &  7149.9    &  484.5   &  17.42  &  0.04   &  $R$  \\
2015.1           &  72.5    &  15.26  &  0.05  &  $R$  &  &  8362.6    &  727.0   &  17.65  &  0.04   &  $R$  \\
2375.1           &  63.0    &  15.63  &  0.08  &  $R$  &  &  9818.1    &  969.0   &  17.98  &  0.05   &  $R$  \\
5120.9           &  189.5   &  16.67  &  0.04  &  $R$  &  &  11758.0   &  969.5   &  18.16  &  0.06   &  $R$  \\
5373.4           &  187.0   &  16.83  &  0.04  &  $R$  &  &  13940.7   &  1212.0  &  18.52  &  0.08   &  $R$  \\
5748.5           &  187.0   &  17.10  &  0.05  &  $R$  &  &  -         &  -       &  -      &  -      &  $-$  \\
%1.23mCalarAlto
\\
\multicolumn{11}{c}{1.23-m Calar Alto} \\
\hline
2467.0           &  60.0    &  15.59  &  0.06  &  $R$  &  &  9692.0    &  60.0    &  17.88  &  0.08   &  $R$  \\
2582.0           &  60.0    &  15.64  &  0.06  &  $R$  &  &  10202.0   &  60.0    &  17.91  &  0.05   &  $R$  \\
4198.0           &  60.0    &  16.48  &  0.05  &  $R$  &  &  10712.0   &  60.0    &  18.02  &  0.05   &  $R$  \\
4661.0           &  60.0    &  16.60  &  0.04  &  $R$  &  &  11215.0   &  60.0    &  18.05  &  0.05   &  $R$  \\
5107.0           &  60.0    &  16.73  &  0.05  &  $R$  &  &  11336.0   &  60.0    &  18.10  &  0.05   &  $R$  \\
5559.0           &  60.0    &  16.85  &  0.05  &  $R$  &  &  11835.0   &  60.0    &  18.15  &  0.05   &  $R$  \\
6008.0           &  60.0    &  16.99  &  0.05  &  $R$  &  &  12337.0   &  60.0    &  18.26  &  0.06   &  $R$  \\
6461.0           &  60.0    &  17.05  &  0.05  &  $R$  &  &  12844.0   &  60.0    &  18.26  &  0.07   &  $R$  \\
6909.0           &  60.0    &  17.23  &  0.04  &  $R$  &  &  13353.0   &  60.0    &  18.28  &  0.05   &  $R$  \\
7358.0           &  60.0    &  17.41  &  0.05  &  $R$  &  &  13905.0   &  60.0    &  18.34  &  0.05   &  $R$  \\
7804.0           &  60.0    &  17.49  &  0.05  &  $R$  &  &  14018.0   &  60.0    &  18.41  &  0.05   &  $R$  \\
8670.0           &  60.0    &  17.73  &  0.05  &  $R$  &  &  14128.0   &  60.0    &  18.35  &  0.05   &  $R$  \\
9182.0           &  60.0    &  17.75  &  0.05  &  $R$  &  &  14239.0   &  60.0    &  18.45  &  0.06   &  $R$  \\
%2.2mCalar Alto
\\
\multicolumn{11}{c}{2.2-m Calar Alto} \\
\hline
9475.0           &  100.0   &  17.67  &  0.03  &  $R$  &  &  12388.0   &  500.0   &  18.13  &  0.03   &  $R$  \\
10064.0          &  500.0   &  17.76  &  0.03  &  $R$  &  &  12962.0   &  500.0   &  18.18  &  0.03   &  $R$  \\
10639.0          &  500.0   &  17.88  &  0.03  &  $R$  &  &  13535.0   &  500.0   &  18.26  &  0.03   &  $R$  \\
11214.0          &  500.0   &  17.97  &  0.03  &  $R$  &  &  14109.0   &  500.0   &  18.33  &  0.03   &  $R$  \\
11787.0          &  500.0   &  18.04  &  0.03  &  $R$  &  &  14706.0   &  500.0   &  18.41  &  0.03   &  $R$  \\
15281.0          &  500.0   &  18.49  &  0.03  &  $R$  &  &  -         &  -       &  -      &  -      &  $-$  \\
%AZT-33IK
\\
\multicolumn{11}{c}{1.6-m AZT-33IK} \\
\hline
37988.0          &  3060.0  &  19.97  &  0.08  &  $R$  &  &  62132.0   &  6600.0  &  20.94  &  0.07   &  $R$  \\
%Zeiss-600
\\
\multicolumn{11}{c}{Zeiss-600} \\
\hline
794.0            &  60.0    &  14.47  &  0.04  &  $R$  &  &  2006.2    &  60.0    &  15.26  &  0.06   &  $V$  \\
853.6            &  60.0    &  14.18  &  0.03  &  $R$  &  &  2068.4    &  60.0    &  15.55  &  0.06   &  $V$  \\
918.4            &  60.0    &  14.08  &  0.02  &  $R$  &  &  2128.0    &  60.0    &  15.52  &  0.06   &  $V$  \\
1022.1           &  60.0    &  14.03  &  0.02  &  $R$  &  &  2230.0    &  2.0     &  15.27  &  0.05   &  $R$  \\
1081.7           &  60.0    &  14.00  &  0.02  &  $R$  &  &  2350.9    &  2.0     &  15.46  &  0.05   &  $R$  \\
1142.2           &  60.0    &  14.07  &  0.02  &  $R$  &  &  2742.3    &  2.0     &  15.54  &  0.06   &  $R$  \\
1204.4           &  60.0    &  14.09  &  0.02  &  $R$  &  &  3133.7    &  2.0     &  15.92  &  0.09   &  $R$  \\
1264.0           &  60.0    &  14.34  &  0.02  &  $R$  &  &  3284.9    &  2.0     &  16.06  &  0.08   &  $R$  \\
1323.7           &  60.0    &  14.29  &  0.03  &  $R$  &  &  3436.1    &  2.0     &  16.08  &  0.04   &  $R$  \\
1385.9           &  60.0    &  14.37  &  0.02  &  $R$  &  &  3557.9    &  2.0     &  16.11  &  0.04   &  $R$  \\
1446.3           &  60.0    &  14.48  &  0.03  &  $R$  &  &  3678.9    &  2.0     &  16.04  &  0.06   &  $R$  \\
1506.0           &  60.0    &  14.52  &  0.03  &  $R$  &  &  3799.9    &  2.0     &  16.16  &  0.05   &  $R$  \\
1568.2           &  60.0    &  14.59  &  0.03  &  $R$  &  &  3921.7    &  2.0     &  16.19  &  0.05   &  $R$  \\
1627.8           &  60.0    &  14.58  &  0.03  &  $R$  &  &  4042.7    &  2.0     &  16.19  &  0.05   &  $R$  \\
1688.3           &  60.0    &  14.74  &  0.03  &  $R$  &  &  4163.6    &  2.0     &  16.30  &  0.06   &  $R$  \\
1814.4           &  60.0    &  14.84  &  0.03  &  $R$  &  &  4345.9    &  4.0     &  16.36  &  0.05   &  $R$  \\
1886.1           &  60.0    &  15.16  &  0.09  &  $V$  &  &  4588.7    &  4.0     &  16.44  &  0.05   &  $R$  \\
1945.7           &  60.0    &  15.38  &  0.04  &  $V$  &  &  4830.6    &  4.0     &  16.59  &  0.06   &  $R$  \\
\enddata
%\tablenotetext{*}{indicates diminished $w_E$ for highest energy photon}
\end{deluxetable}

\begin{deluxetable}{rrrrccrrrrc}
 \tablewidth{0pt}
 \tablecaption{Photomertric observations for GRB 110205A from $Swift$/UVOT\label{tab:UVOT_photometry_data}}
  \tablehead{\colhead{T-T$_0$ (s)} & \colhead{Error (s)} & \colhead{Count} & \colhead{Error} & \colhead{Filter} & \colhead{} & \colhead{T-T$_0$ (s)} & \colhead{Error (s)} & \colhead{Count} & \colhead{Error} & \colhead{Filter}}
\startdata
169.0    &  5.0    &  3.905   &  0.890  &  $white$  &  &   24517.0   &  149.9  &  1.133  &  0.130  &  $white$ \\
179.0    &  5.0    &  2.807   &  0.812  &  $white$  &  &   24734.3   &  63.5   &  1.037  &  0.198  &  $white$ \\
189.0    &  5.0    &  2.963   &  0.840  &  $white$  &  &   35675.8   &  149.9  &  0.654  &  0.122  &  $white$ \\
199.0    &  5.0    &  3.326   &  0.877  &  $white$  &  &   35979.3   &  149.9  &  0.653  &  0.123  &  $white$ \\
209.0    &  5.0    &  9.487   &  1.203  &  $white$  &  &   36284.1   &  149.9  &  0.474  &  0.089  &  $white$ \\
219.0    &  5.0    &  7.664   &  1.120  &  $white$  &  &   75564.8   &  149.9  &  0.234  &  0.080  &  $white$ \\
229.0    &  5.0    &  4.538   &  0.939  &  $white$  &  &   75868.3   &  149.9  &  0.256  &  0.081  &  $white$ \\
239.0    &  5.0    &  4.255   &  0.916  &  $white$  &  &   76171.9   &  149.9  &  0.316  &  0.083  &  $white$ \\
249.0    &  5.0    &  2.964   &  0.845  &  $white$  &  &   87890.9   &  92.9   &  0.177  &  0.099  &  $white$ \\
259.0    &  5.0    &  2.689   &  0.831  &  $white$  &  &   99845.1   &  88.9   &  0.085  &  0.100  &  $white$ \\
269.0    &  5.0    &  2.722   &  0.829  &  $white$  &  &   110704.6  &  322.4  &  0.106  &  0.052  &  $white$ \\
279.0    &  5.0    &  2.633   &  0.802  &  $white$  &  &   117007.3  &  147.4  &  0.008  &  0.075  &  $white$ \\
289.0    &  5.0    &  3.229   &  0.858  &  $white$  &  &   122512.3  &  237.9  &  0.028  &  0.059  &  $white$ \\
299.0    &  5.0    &  3.095   &  0.863  &  $white$  &  &   128037.7  &  322.4  &  0.097  &  0.052  &  $white$ \\
308.9    &  4.9    &  2.720   &  0.826  &  $white$  &  &   133882.8  &  299.9  &  0.100  &  0.054  &  $white$ \\
611.8    &  9.9    &  20.129  &  1.184  &  $white$  &  &   139643.2  &  305.9  &  0.034  &  0.052  &  $white$ \\
784.5    &  9.9    &  63.173  &  2.257  &  $white$  &  &   145379.3  &  319.4  &  0.047  &  0.051  &  $white$ \\
950.3    &  74.8   &  88.866  &  1.043  &  $white$  &  &   151192.4  &  308.4  &  0.167  &  0.055  &  $white$ \\
1190.9   &  9.9    &  95.569  &  3.037  &  $white$  &  &   156929.3  &  321.9  &  0.038  &  0.051  &  $white$ \\
1365.9   &  9.9    &  78.130  &  2.615  &  $white$  &  &   162706.6  &  321.4  &  0.041  &  0.051  &  $white$ \\
1538.4   &  9.9    &  64.515  &  2.290  &  $white$  &  &   168487.5  &  320.9  &  0.042  &  0.051  &  $white$ \\
1711.7   &  9.9    &  52.993  &  2.027  &  $white$  &  &   185834.0  &  315.4  &  0.006  &  0.051  &  $white$ \\
5166.4   &  99.9   &  10.865  &  0.283  &  $white$  &  &   203156.9  &  319.9  &  0.088  &  0.052  &  $white$ \\
6602.8   &  99.9   &  7.511   &  0.246  &  $white$  &  &   209526.6  &  120.9  &  0.019  &  0.083  &  $white$ \\
11906.7  &  149.9  &  2.834   &  0.152  &  $white$  &  &   220477.0  &  322.9  &  0.022  &  0.050  &  $white$ \\
12210.4  &  149.9  &  2.752   &  0.150  &  $white$  &  &   232055.5  &  315.9  &  0.042  &  0.052  &  $white$ \\
12513.8  &  149.9  &  2.755   &  0.151  &  $white$  &  &   -         &  -      &  -      &  -      &  $white$ \\
\\
150.5    &  4.9    &  1.107   &  0.449  &  $v$      &  &   52773.8   &  149.9  &  0.077  &  0.039  &  $v$     \\
661.9    &  9.9    &  6.853   &  0.643  &  $v$      &  &   53077.3   &  149.9  &  0.067  &  0.038  &  $v$     \\
834.2    &  9.9    &  13.718  &  0.903  &  $v$      &  &   71192.4   &  18.0   &  0.069  &  0.121  &  $v$     \\
1066.4   &  9.9    &  19.918  &  1.096  &  $v$      &  &   80344.4   &  149.9  &  0.004  &  0.035  &  $v$     \\
1241.8   &  9.9    &  18.517  &  1.053  &  $v$      &  &   80648.3   &  149.9  &  0.047  &  0.037  &  $v$     \\
1415.6   &  9.9    &  11.828  &  0.837  &  $v$      &  &   111203.5  &  169.5  &  0.063  &  0.036  &  $v$     \\
1588.3   &  9.9    &  11.398  &  0.820  &  $v$      &  &   122914.8  &  158.1  &  0.038  &  0.035  &  $v$     \\
1761.6   &  9.9    &  10.388  &  0.788  &  $v$      &  &   128677.6  &  310.6  &  0.049  &  0.026  &  $v$     \\
5576.8   &  99.9   &  1.641   &  0.115  &  $v$      &  &   134477.4  &  288.0  &  0.013  &  0.026  &  $v$     \\
7013.2   &  99.9   &  1.123   &  0.101  &  $v$      &  &   157567.5  &  309.5  &  0.018  &  0.025  &  $v$     \\
16773.0  &  149.9  &  0.338   &  0.051  &  $v$      &  &   163345.1  &  309.6  &  0.015  &  0.024  &  $v$     \\
17076.6  &  149.9  &  0.302   &  0.050  &  $v$      &  &   169123.9  &  308.7  &  0.018  &  0.025  &  $v$     \\
17380.0  &  149.9  &  0.337   &  0.051  &  $v$      &  &   192030.6  &  102.3  &  0.039  &  0.045  &  $v$     \\
41428.4  &  149.9  &  0.048   &  0.037  &  $v$      &  &   203748.6  &  264.3  &  0.050  &  0.028  &  $v$     \\
41731.9  &  149.9  &  0.082   &  0.039  &  $v$      &  &   221119.4  &  312.7  &  0.016  &  0.024  &  $v$     \\
42035.4  &  149.9  &  0.126   &  0.041  &  $v$      &  &   232682.9  &  304.9  &  0.013  &  0.025  &  $v$     \\
52470.2  &  149.9  &  0.053   &  0.037  &  $v$      &  &   -         &  -      &  -      &  -      &  $v$     \\
\\
587.4    &  9.9    &  6.547   &  0.659  &  $b$      &  &   75259.9   &  149.9  &  0.114  &  0.050  &  $b$     \\
760.2    &  9.9    &  20.774  &  1.144  &  $b$      &  &   87035.9   &  149.9  &  0.116  &  0.050  &  $b$     \\
1164.5   &  9.9    &  33.987  &  1.501  &  $b$      &  &   87339.4   &  149.9  &  0.052  &  0.048  &  $b$     \\
1341.7   &  9.9    &  28.043  &  1.347  &  $b$      &  &   87642.9   &  149.9  &  0.034  &  0.047  &  $b$     \\
1514.1   &  9.9    &  21.021  &  1.154  &  $b$      &  &   99662.9   &  88.9   &  0.012  &  0.060  &  $b$     \\
1686.7   &  9.9    &  18.354  &  1.076  &  $b$      &  &   110054.4  &  322.4  &  0.075  &  0.034  &  $b$     \\
1860.6   &  9.9    &  15.713  &  1.035  &  $b$      &  &   116707.9  &  147.4  &  0.024  &  0.047  &  $b$     \\
6397.2   &  99.9   &  2.714   &  0.148  &  $b$      &  &   122031.6  &  237.9  &  0.098  &  0.040  &  $b$     \\
10993.9  &  149.9  &  0.958   &  0.093  &  $b$      &  &   127387.6  &  322.4  &  0.051  &  0.032  &  $b$     \\
11297.6  &  149.9  &  1.020   &  0.094  &  $b$      &  &   133277.3  &  299.9  &  0.056  &  0.033  &  $b$     \\
11601.7  &  149.9  &  1.102   &  0.095  &  $b$      &  &   139026.1  &  305.9  &  0.038  &  0.033  &  $b$     \\
23604.9  &  149.9  &  0.332   &  0.059  &  $b$      &  &   144734.8  &  319.4  &  0.059  &  0.033  &  $b$     \\
23908.6  &  149.9  &  0.405   &  0.061  &  $b$      &  &   150570.4  &  308.4  &  0.026  &  0.033  &  $b$     \\
24212.1  &  149.9  &  0.359   &  0.060  &  $b$      &  &   156280.2  &  321.9  &  0.057  &  0.033  &  $b$     \\
30146.5  &  149.9  &  0.265   &  0.056  &  $b$      &  &   162058.5  &  321.4  &  0.047  &  0.033  &  $b$     \\
30449.9  &  149.9  &  0.308   &  0.058  &  $b$      &  &   167840.3  &  320.9  &  0.019  &  0.032  &  $b$     \\
30676.6  &  73.1   &  0.192   &  0.085  &  $b$      &  &   190956.6  &  319.9  &  0.086  &  0.034  &  $b$     \\
34763.7  &  149.9  &  0.194   &  0.054  &  $b$      &  &   196608.1  &  144.9  &  0.009  &  0.046  &  $b$     \\
35067.2  &  149.9  &  0.063   &  0.049  &  $b$      &  &   202511.7  &  319.9  &  0.003  &  0.031  &  $b$     \\
35370.8  &  149.9  &  0.167   &  0.053  &  $b$      &  &   209280.3  &  120.9  &  0.045  &  0.053  &  $b$     \\
59136.9  &  149.9  &  0.053   &  0.047  &  $b$      &  &   219825.9  &  322.9  &  0.006  &  0.032  &  $b$     \\
59440.3  &  149.9  &  0.106   &  0.050  &  $b$      &  &   225640.3  &  315.9  &  0.040  &  0.032  &  $b$     \\
74652.2  &  149.9  &  0.012   &  0.045  &  $b$      &  &   237196.2  &  315.9  &  0.008  &  0.032  &  $b$     \\
\\
336.9    &  15.0   &  0.311   &  0.170  &  $u$      &  &   65363.9   &  68.7   &  0.067  &  0.055  &  $u$     \\
366.9    &  15.0   &  0.263   &  0.175  &  $u$      &  &   86123.4   &  149.9  &  0.047  &  0.032  &  $u$     \\
396.9    &  15.0   &  0.453   &  0.187  &  $u$      &  &   86730.7   &  149.9  &  0.027  &  0.031  &  $u$     \\
426.9    &  15.0   &  0.843   &  0.223  &  $u$      &  &   109403.9  &  322.4  &  0.029  &  0.021  &  $u$     \\
456.9    &  15.0   &  0.618   &  0.199  &  $u$      &  &   114916.8  &  56.3   &  0.029  &  0.051  &  $u$     \\
486.9    &  15.0   &  1.113   &  0.239  &  $u$      &  &   116407.8  &  147.4  &  0.031  &  0.031  &  $u$     \\
516.9    &  15.0   &  1.753   &  0.288  &  $u$      &  &   126736.9  &  322.4  &  0.037  &  0.021  &  $u$     \\
546.9    &  15.0   &  2.956   &  0.352  &  $u$      &  &   138408.5  &  305.9  &  0.002  &  0.020  &  $u$     \\
566.8    &  4.9    &  3.160   &  0.632  &  $u$      &  &   144090.4  &  319.4  &  0.002  &  0.020  &  $u$     \\
735.3    &  9.9    &  12.198  &  0.843  &  $u$      &  &   149947.9  &  308.4  &  0.030  &  0.022  &  $u$     \\
1140.4   &  9.9    &  21.735  &  1.144  &  $u$      &  &   161409.7  &  321.4  &  0.028  &  0.021  &  $u$     \\
1315.2   &  9.9    &  18.643  &  1.052  &  $u$      &  &   167193.0  &  320.9  &  0.003  &  0.020  &  $u$     \\
1489.0   &  9.9    &  15.334  &  0.949  &  $u$      &  &   184561.2  &  315.4  &  0.047  &  0.022  &  $u$     \\
1661.8   &  9.9    &  13.044  &  0.871  &  $u$      &  &   190310.9  &  319.9  &  0.029  &  0.022  &  $u$     \\
1835.9   &  9.9    &  11.387  &  0.833  &  $u$      &  &   196145.9  &  311.4  &  0.003  &  0.020  &  $u$     \\
6192.1   &  99.9   &  1.905   &  0.114  &  $u$      &  &   201865.8  &  319.9  &  0.009  &  0.020  &  $u$     \\
7590.9   &  62.7   &  1.557   &  0.146  &  $u$      &  &   209033.6  &  120.9  &  0.008  &  0.033  &  $u$     \\
22692.2  &  149.9  &  0.340   &  0.048  &  $u$      &  &   213551.4  &  300.9  &  0.009  &  0.021  &  $u$     \\
22995.8  &  149.9  &  0.251   &  0.044  &  $u$      &  &   219174.3  &  322.9  &  0.033  &  0.021  &  $u$     \\
23299.6  &  149.9  &  0.267   &  0.044  &  $u$      &  &   225002.7  &  315.9  &  0.004  &  0.020  &  $u$     \\
29233.0  &  149.9  &  0.174   &  0.039  &  $u$      &  &   230780.7  &  315.9  &  0.029  &  0.021  &  $u$     \\
29536.8  &  149.9  &  0.163   &  0.039  &  $u$      &  &   236558.5  &  315.9  &  0.017  &  0.021  &  $u$     \\
29841.2  &  149.9  &  0.101   &  0.035  &  $u$      &  &   254513.9  &  859.1  &  0.005  &  0.012  &  $u$     \\
47582.1  &  149.9  &  0.080   &  0.034  &  $u$      &  &   260333.6  &  859.3  &  0.018  &  0.012  &  $u$     \\
47885.6  &  149.9  &  0.095   &  0.036  &  $u$      &  &   266093.7  &  859.3  &  0.012  &  0.012  &  $u$     \\
48069.1  &  29.9   &  0.277   &  0.111  &  $u$      &  &   271853.7  &  859.2  &  0.030  &  0.013  &  $u$     \\
58224.4  &  149.9  &  0.068   &  0.033  &  $u$      &  &   277673.7  &  859.3  &  0.022  &  0.013  &  $u$     \\
58527.9  &  149.9  &  0.063   &  0.033  &  $u$      &  &   282893.6  &  319.2  &  0.002  &  0.020  &  $u$     \\
58831.6  &  149.9  &  0.112   &  0.036  &  $u$      &  &   317849.0  &  104.0  &  0.027  &  0.037  &  $u$     \\
\\
710.9    &  9.9    &  0.481   &  0.174  &  $uvw1$   &  &   5987.1    &  99.9   &  0.110  &  0.030  &  $uvw1$  \\
1115.9   &  9.9    &  1.735   &  0.309  &  $uvw1$   &  &   7423.1    &  99.9   &  0.107  &  0.031  &  $uvw1$  \\
1290.9   &  9.9    &  0.869   &  0.226  &  $uvw1$   &  &   18825.6   &  383.8  &  0.024  &  0.011  &  $uvw1$  \\
1464.7   &  9.9    &  0.452   &  0.173  &  $uvw1$   &  &   28626.8   &  449.9  &  0.013  &  0.008  &  $uvw1$  \\
1637.4   &  9.9    &  0.918   &  0.235  &  $uvw1$   &  &   46975.9   &  449.9  &  0.007  &  0.008  &  $uvw1$  \\
1811.6   &  9.9    &  0.525   &  0.183  &  $uvw1$   &  &   57618.1   &  449.9  &  0.010  &  0.008  &  $uvw1$  \\
\enddata
\end{deluxetable}


\begin{thebibliography}{}

\bibitem[{{Abdo} {et~al.}(2009c)}]{abdo09} {Abdo}, A.~A., et al., 2009, Nature, 462, 331 %090510

\bibitem[{{Akerlof} {et~al.}(1999)}]{akerlof99} {Akerlof}, C.~W., et al., 1999, Nature, 398, 400

\bibitem[{{Akerlof} \& {Swan}(2007)}]{akerlof07} {Akerlof}, C.~W. \& {Swan}, H.~F., 2007, ApJ, 671, 1868

\bibitem[{{Andreev} {et~al.}(2011)}] {andreev11} {Andreev}, M., Sergeev, A., Pozanenko, A., 2011, GCN Circ., 11641

\bibitem[{{Atwood} {et~al.}(2009)}]{atwood09} {Atwood}, W.~B., et~al., 2009, ApJ, 697, 1071

\bibitem[{{Band} {et~al.}(1993)}] {band93} {Band}, D., et~al. 1993, ApJ, 413, 281

\bibitem[{{Baltz} {et~al.}(2005)}] {baltz05} {Baltz}, E. A. \& Hui, L., 2005, ApJ, 618, 403

\bibitem[{{Barthelmy} {et~al.}(2005)}] {barthelmy93} {Barthelmy}, S., et al. 2005, Space Sci. Rev., 120, 143

\bibitem[{{Beardmore} {et~al.}(2011)}] {beardmore11} {Beardmore}, A., et al. 2011, GCN Circ., 11629

\bibitem{} Beloborodov, A. M., 2010, MNRAS, 407, 1033

\bibitem[{{Blandford} {et~al.}(1976)}] {blandford76} Blandford, R. D. \& McKee, C. F., Physics of Fluids, 1976, 19, 1130

\bibitem[{{Breeveld} {et~al.}(2010)}] {breeveld10} {Breeveld}, A. A., et~al., 2010, MNRAS, 406, 1687

\bibitem[{{Breeveld} {et~al.}(2011)}] {breeveld11} {Breeveld}, A. A., et~al., 2011, AIP Conference Proceedings, 1358, 373

\bibitem[{{Burrows} {et~al.}(2005{\natexlab{a}}){Burrows}, {Hill}, {Nousek},
  {Kennea}, {Wells}, {Osborne}, {Abbey}, {Beardmore}, {Mukerjee}, {Short},
  {Chincarini}, {Campana}, {Citterio}, {Moretti}, {Pagani}, {Tagliaferri},
  {Giommi}, {Capalbi}, {Tamburelli}, {Angelini}, {Cusumano}, {Br{\"a}uninger},
  {Burkert}, \& {Hartner}}]{Burrows2005a}
{Burrows}, D.~N., {Hill}, J.~E., {Nousek}, J.~A., {et~al.} 2005a,
  Space Sci. Rev., 120, 165

\bibitem{} Burrows, D. N. et al. 2005b, Science, 309, 1833

\bibitem[{{Cenko} {et~al.}(2006)}] {cenko06} {Cenko}, S.~B., et~al., 2006, ApJ, 652, 490

\bibitem[{{Cenko} {et~al.}(2011)}] {cenko11} {Cenko}, S.~B., Hora, J. \& Bloom, S. 2011, GCN Circ., 11638

\bibitem[{{Chester} \& {Beardmore}(2011)}] {chester11} {Chester}, M. M., \& {Beardmore}, A.~P. 2011, GCN Circ., 11634

\bibitem{} Chincarini, G. et al. 2007, ApJ, 671, 1903

\bibitem[{{Cucchiara} {et~al.}(2011)}] {cucchiara11} {Cucchiara}, A., et al. 2011, arXiv:1107.3352

\bibitem[{{Corsi} {et~al.}(2005)}] {corsi05} {Corsi}, A., et~al., 2005, A\&A, 438, 829

\bibitem{} Covino, S. et al. 1999, A\&A, 348, L1

\bibitem{} Covino, S. et al. 2003, A\&A, 400, L9

\bibitem[{{Dai} (2003)}] {dai03} {Dai}, Z. G. \& Wu, X. F., 2003, ApJ, 591, L21

\bibitem{} Daigne, F. \& Mochkovitch, R. 1998, MNRAS, 296, 275

\bibitem[{{De Pasquale } {et~al.}(2003)}] {pasquale03} {De Pasquale}, M. et al., 2003, ApJ, 592, 1018

\bibitem[{{De Pasquale } {et~al.}(2010)}] {pasquale10} {De Pasquale}, M. et al., 2010, ApJ, 709, 146

\bibitem[{{Efimov} {et~al.}(2003)}] {efimov03} {Efimov}, Y. et al., 2003, GCN Circ., 2144

\bibitem[{{Evans} {et~al.}(2007){Evans}, {Beardmore}, {Page}, {Tyler},
  {Osborne}, {Goad}, {O'Brien}, {Vetere}, {Racusin}, {Morris}, {Burrows},
  {Capalbi}, {Perri}, {Gehrels}, \& {Romano}}]{Evans2007a}
{Evans}, P.~A., {Beardmore}, A.~P., {Page}, K.~L., {et~al.} 2007,
\aap, 469, 379

\bibitem[{{Evans} {et~al.}(2009){Evans}, {Beardmore}, {Page}, {Osborne},
  {O'Brien}, {Willingale}, {Starling}, {Burrows}, {Godet}, {Vetere}, {Racusin},
  {Goad}, {Wiersema}, {Angelini}, {Capalbi}, {Chincarini}, {Gehrels}, {Kennea},
  {Margutti}, {Morris}, {Mountford}, {Pagani}, {Perri}, {Romano}, \&
  {Tanvir}}]{Evans2009a}
{Evans}, P.~A., {Beardmore}, A.~P., {Page}, K.~L., {et~al.} 2009,
\mnras, 397,
  1177

\bibitem[{{Evans} {et~al.}(2010)}] {evans10} {Evans}, P. A. et al., 2010, \aap, 519, 102

\bibitem{} Fan, Y.-Z., Dai, Z.-G., Huang, Y.-F., Lu, T. 2002, ChJAA, 2, 449

\bibitem[{{Fan} {et~al.}(2005)}] {fan05} {Fan}, Y. \& Wei, D., 2005, MNRAS, 364, L42

\bibitem{} Fan, Y. Z., Zhang, B., Wei, D. M., 2009, Physical Review D, 79, 021301

\bibitem[{{Fruchter} {et~al.}(2001)}] {fruchter01} {Fruchter}, A., Krolik, J. H., \& Rhoads, J. E., 2001, ApJ, 563, 597

\bibitem[{{Galama} (1998)}] {galama98} {Galama}, T. J., 1998, ApJ, 497, L13

\bibitem[{{Gao} (2011)}] {gao11} {Gao}, W., 2011, arXiv:1104.3382

\bibitem[{{Garnavich} (2000)}] {garnavich00} {Garnavich}, P. M., Loeb, A. \& Stanek, K. Z., 2000, ApJ, 544, L11

\bibitem[{{Gaudi} (2001)}] {gaudi01} {Gaudi}, B. S., Granot, J. \& Loeb, A., 2001, ApJ, 561, 178

\bibitem[{{Gehrels} {et~al.}(2004)}] {gehrels04} {Gehrels}, N., et al., 2004, ApJ, 611, 1005

\bibitem[{{Gendre} {et~al.}(2011)}] {gendre11} {Gendre}, B., et al., 2011, arXiv:1110.0734

\bibitem{} Giannios D., 2008, A\&A, 480, 305

\bibitem{} Ghisellini, G., Celotti, A., Lazzati, D. 2000, MNRAS, 313, L1

\bibitem{} Ghisellini, G. \& Lazzati, D. 1999, ApJ, 309, 7

\bibitem[{{Golenetskii} {et~al.}(2011)}] {golenetskii11} {Golenetskii}, S., et al., 2011, GCN Circ., 11659

\bibitem[{{Gorosabel} {et~al.}(2011)}] {gorosabel11} {Gorosabel}, J., Duffard, R., Kubanek, P. \& Guijarro, A., 2011, GCN Circ., 11696

\bibitem[{{Granot} (2002)}] {granot02} {Granot}, J. \& Sari, R., 2002, ApJ, 568, 820

\bibitem{} Greiner, J. et al. 2003, Nature, 426, 157

\bibitem{} Gruzinov, A., Waxman, E. 1999, ApJ, 511, 852

\bibitem{} Gruber, D., et al., 2011, A\&A, 528, 15

\bibitem[{{Hentunen} {et~al.}(2011)}] {hentunen11} {Hentunen}, V. P., Nissinen, M., Salmi, T., 2011, GCN Circ., 11637

\bibitem[{{Im} {et~al.}(2011)}] {im11} {Im}, M. \& Urata., Y., 2011, GCN Circ., 11643

\bibitem[{{Jin} {et~al.}(2007)}] {jin07} {Jin}, Z. \& Fan., Y., 2007, MNRAS, 378, 1043

\bibitem[Jung et al.(1989)]{jung1989} Jung, G.V., 1989, ApJ, 338, 972

\bibitem[{{Kalberla} \& {Kalberla}(2005)}] {kalberla05} {Kalberla}, P. M., Burton, W. B., Hartmann, D., Arnal, E. M., Bajaja, E., Morras, R., Poppel, W. G., 2005, A\&A, 440, 775

\bibitem[{{Klebesadel} \& {Klebesadel}(2009)}] {klebesadel73} {Klebesadel}, R., Strong, I. \& Olson, R. 1973, ApJ, 182, L85

\bibitem[{{Klotz} {et~al.}(2011a)}] {klotz11a} {Klotz}, A., et al., 2011a, GCN Circ., 11630

\bibitem[{{Klotz} {et~al.}(2011b)}] {klotz11b} {Klotz}, A., et al., 2011b, GCN Circ., 11632

\bibitem[{{Kobayashi}(2000)}] {kobayashi00} {Kobayashi}, S., 2000, ApJ, 545, 807

\bibitem[{{Kobayashi} {et al.}(2007)}] {kobayashi07} {Kobayashi}, S.,
Zhang, B., M\'esz\'aros, P., Burrows, D. N.  2007, ApJ, 655, 391

\bibitem[{{Kong}(2010)}] {kong10} {Kong}, S. W., 2010, MNRAS, 402, 409

\bibitem[{{Kumar} {et al.}(2008)}] {kumar08} {Kumar}, P., \& {McMahon}, R. 2008, MNRAS, 384, 33

\bibitem{} Kumar, P., Panaitescu, A. 2000, ApJ, 541, L51

\bibitem{} Kumar, P., Panaitescu, A. 2008, MNRAS, 391, L19

\bibitem[{{Lazzati} (2002)}] {lazzati02} {Lazzati}, D., et al., 2002a, MNRAS, 330, 583

\bibitem[{{Lazzati} (2002)}] {lazzati02} {Lazzati}, D., et al., 2002b, A\&A, 396, L5

\bibitem{} Lazzati, D., Morsony, B. J., Begelman, M. C., 2009, ApJ, 700, 47

\bibitem{} Lazzati, D., \& Begelman, M. C., 2010, ApJ, 725, 1137

\bibitem{} Lazzati, D., Morsony, B. J., Begelman, M. C., 2011, ApJ, 732, 34

\bibitem{} Lee, I., Im, M., \& Urata, Y. 2010, JKAS 43, 95

\bibitem[{{Liang} (2006)}] {liang06} {Liang}, E., et al., 2006, ApJ, 646, 351

\bibitem[{{Liang} (2010)}] {liang10} {Liang}, E., Yi, S., Zhang, J., Lu, H., Zhang, B.-B. \& Zhang, B., 2010, ApJ, 725, 2209

\bibitem{} Li, Z., Waxman, E., 2008, ApJ, 674, L65

\bibitem{} Mao, J., Malesani, D., D'Avanzo, P., et al. 2011, arXiv:1112.0744

\bibitem[{{Meegan} {et~al.}(2009)}] {meegan09} {Meegan}, C., et al., 2009, ApJ, 702, 791

\bibitem{} Medvedev, M. V. 2006, ApJ, 637, 869

\bibitem{} Medvedev, M. V., Loeb, A. 1999, ApJ, 526, 697

\bibitem[]{meszaros06} M\'esz\'aros, P. 2006, Rep. Prog. Phys. 69, 2259

\bibitem[]{meszaros97} M\'esz\'aros, P. \& Rees, M. J. 1997, ApJ, 476, 232

\bibitem[]{meszaros99} M\'esz\'aros, P. \& Rees, M. J. 1999, MNRAS, 306, L39

\bibitem[]{meszaros06} M\'esz\'aros, P., et al., 2006, Rep. Prog. Phys., 69, 2259

\bibitem[]{molinari07} Molinari, E., et al., 2007, A\&A, 469, 13

\bibitem[{{Nakar} (2007)}] {nakar07} {Nakar}, E. \& Granot, J., 2007, MNRAS, 380, 1744

\bibitem[]{nardini} Nardini, M., et al., 2011, A\&A, 531, 39

\bibitem{} Nishikawa, K.-I. et al. 2009, ApJ, 698, L10

\bibitem[{{Oates} {et~al.}(2009)}] {oates09b} Oates S. R. et al., 2009, MNRAS, 395, 490

\bibitem[{{Ovaldsen} (2007)}] {ovaldsen07} {Ovaldsen}, J., et al., 2007, ApJ, 662, 294

\bibitem[]{Palshin} Pal'shin, V. 2011, GCN Circ., 11697

\bibitem{} Panaitescu A., Kumar P., 2004, MNRAS, 353, 511

\bibitem[{{Panaitescu} \& {Vestrand}(2011)}] {panaitescu08} {Panaitescu}, A. \& {Vestrand}, W.~T., 2011, arXiv:1009.3947

\bibitem{} Pe'er, A., Waxman, E., 2005, ApJ, 633, 1018

\bibitem{} Pe'er, A., M\'esz\'aros, P., Rees, M. J., 2006, ApJ, 642, 995

\bibitem{} Pe'er, A., Zhang, B. 2006, ApJ, 653, 454

\bibitem{} Piran, T., 2005, AIP Conference Proceedings, 784, 164

\bibitem{} Piran, T., Sari, R., Zou, Y. C., 2009, MNRAS, 393, 1107

\bibitem[{{Poole} {et~al.}(2008)}] {poole08} Poole G. et al.  2008, MNRAS, 383, 627

\bibitem[{{Pozanenko} {et~al.}(2008)}] {pozanenko08} Pozanenko, A., et al., 2008, Astronomy Letters, 34, 141–144

\bibitem[{{Quimby} {et~al.}(2006)}] {quimby06} {Quimby}, R.~M., et~al. 2006, ApJ, 640, 402

\bibitem[]{Racusin} Racusin, J. et al., Nature, 455, 183

\bibitem[]{Rees92} Rees, M. J. \& M\'esz\'aros, P. 1992, MNRAS, 258, 41

\bibitem[]{Rees94} Rees, M. J. \& M\'esz\'aros, P. 1994, ApJ, 430, L93

\bibitem[{{Roming} {et~al.}(2005){Roming}, {Kennedy}, {Mason}, {Nousek}, {Ahr},
  {Bingham}, {Broos}, {Carter}, {Hancock}, {Huckle}, {Hunsberger}, {Kawakami},
  {Killough}, {Koch}, {McLelland}, {Smith}, {Smith}, {Soto}, {Boyd},
  {Breeveld}, {Holland}, {Ivanushkina}, {Pryzby}, {Still}, \&
  {Stock}}]{Roming2005}
{Roming}, P.~W.~A., {Kennedy}, T.~E., {Mason}, K.~O., {et~al.} 2005,
Space Sci.
  Rev., 120, 95

\bibitem[{{Rossi} {et~al.}(2011)}] {rossi11} {Rossi}, A., et al. 2011, A\&A, 529, 142

\bibitem[Rothschild et al.(1998)]{rothschild1998} Rothschild, R.E., et al. 1998, ApJ, 496, 538

\bibitem[{{Rykoff} {et~al.}(2009)}] {rykoff09} {Rykoff}, E. S., et al. 2009, ApJ, 702, 489

\bibitem[{{Sahu} {et~al.}(2011)}] {sahu11} {Sahu}, D. K., \& Anto, P., 2011, GCN Circ., 11670

\bibitem[Sakamoto et al.(2011a)]{sakamoto2011a} Sakamoto, T., et al. 2011a, ApJS, 195, 2

\bibitem[Sakamoto et al.(2011b)]{sakamoto2011b} Sakamoto, T., et al. 2011b, PASJ, 63, 215

\bibitem[{{Sakamoto} {et~al.}(2011)}] {sakamoto11c} {Sakamoto}, T., et al., 2011c, GCN Circ., 11692

\bibitem[{{Sari} (1999)}] {sari99} {Sari}, R., 1999, ApJ, 524, 43

\bibitem[{{Sari} \& {Esin}(2001)}] {sari01} {Sari}, R. \& {Esin}, A.~A., 2001, ApJ, 548, 787

\bibitem[{{Sari} \& {Piran}(1995)}] {sari95} {Sari}, R. \& {Piran}, T., 1995, ApJ, 455, L143

\bibitem[{{Sari} \& {Piran}(1999a)}] {sari99a} {Sari}, R. \& {Piran}, T., 1999, ApJ, 517, L109

\bibitem[{{Sari}, {Piran} \& {Narayan}(1998)}] {sari98} {Sari}, R., {Piran}, T. \& {Narayan} R., 1998, ApJ, 497, L17

\bibitem[{{Sari} (1999)}] {sari99} {Sari}, R., {Piran}, T. \& {Halpern} J., 1999, ApJ, 519, L17

\bibitem[{{Savaglio} (2009)}] {savaglio09} {Savaglio}, S., Glazebrook, K. \& LeBorgne, D., 2009, ApJ, 691, 182

\bibitem[{{Schaefer} {et~al.}(2011)}] {schaefer11} {Schaefer}, B. E., et al., 2011, GCN Circ., 11631

\bibitem[{{Schlegel} {et~al.}(1998)}] {schlegel98} {Schlegel}, D.~J., {Finkbeiner}, D.~P. \& {Davis}, M.  1998, ApJ, 500, 525

\bibitem{} Shen, R., Kumar P., Piran T., 2010, MNRAS, 403, 229

\bibitem{} Shen, R. \& Matzner C., 2012, ApJ, 744, 36

\bibitem[{Shen}] {shen09} {Shen}, R. \& Zhang, B., 2009, MNRAS, 398, 1936

\bibitem[{{Silva} {et~al.}(2011)}] {silva11} {Silva}, R., Fumagalli., M., Worseck., G. \& Prochaska, X. 2011, GCN Circ., 11635

\bibitem[Sizun et al.(2004)]{sizun2004} Sizun, P. et al. 2004, in Proceedings of the 5th INTEGRAL Workshop on the INTEGRAL Universe, ed. V. Sch\"{o}nfelder, G. Lichti \& C. Winkler, ESA SP-552, 815

\bibitem[{{Smith} {et~al.}(2002)}] {smith98} {Smith}, J., et al. 2002, AJ, 123, 2121

\bibitem[{{Sugita} {et~al.}(2011)}] {sugita11} {Sugita}, S., et al., 2011, GCN Circ., 11682

\bibitem{} Tagliaferri, G. et al. 2005, Nature, 436, 985

\bibitem{} Toma, K., Wu, X.-F., M\'esz\'aros, P., 2011, MNRAS, 415, 1663

\bibitem[{{Ukwatta} {et~al.}(2011)}] {ukwatta11} {Ukwatta}, T., et al., 2011, GCN Circ., 11655

\bibitem[{{Updike} {et~al.}(2008)}] {updike08} {Updike}, A.~C., et al., 2008, ApJ, 685, 361

\bibitem[{{Urata} {et~al.}(2011)}] {urata11} {Urata}, Y., Chuang, C. \& Huang, K., 2011, GCN Circ., 11648

\bibitem{} van Eerten, H. J., MacFadyen, A. I., 2011, ApJ, 733, 37

\bibitem[{{Volnova} {et~al.}(2010)}] {volnova11} {Volnova}, A., et al., 2010, GCN Circ., 11270

\bibitem[{{Volnova} {et~al.}(2011)}] {volnova11} {Volnova}, A., Klunko, E., Pozanenko, A., 2011, GCN Circ., 11672

\bibitem[{{Vreeswijk} {et~al.}(2011)}] {vreeswijk11} {Vreeswijk}, P., et al., 2011, GCN Circ., 11640

\bibitem{} Vurm, I., Beloborodov, A. M., Poutanen, J. 2011, ApJ, in press (arXiv:1104.0394)

\bibitem[{{Waxman} (2000)}] {waxman00} {Waxman}, E. \& Draine, B. T., 2000, ApJ, 537, 796

\bibitem{} Weibel, E. S. 1959, PRL, 2, 83

\bibitem[{{Wijers} {et~al.}(1997)}] {wijers97} {Wijers}, R., Rees, M., M\'esz\'aros, P., 1997, MNRAS, 288, 51

\bibitem[{{Xue} {et~al.}(2009)}] {xue09} {Xue}, R., Fan, Y. \& Wei, D., 2009, A\&A, 498, 671

\bibitem[{{Yamaoka} {et~al.}(2009)}] {yamaoka09} {Yamaoka}, K., et al., 2009, PASJ, 61, 35

\bibitem[{{Yu} {et~al.}(2009)}] {yu09} Yu, Y. W., Wang, X. Y. \& Dai, Z. G., 2009, ApJ, 692, 1662

\bibitem[{{Zhang}, (2011)}] {zhang11} {Zhang}, B., 2011, Comptes Rendus Physique, 12, 206
(arXiv:1104.0932)

\bibitem[{{Zhang}, {et al.}(2006)}] {zhang06} {Zhang}, B., Fan, Y. Z., Dyks, J. et al.
2006, ApJ, 642, 354

\bibitem[{{Zhang}, {Kobayashi}\& { Me\'sza\'ros}(2003)}] {zhang03} {Zhang}, B., {Kobayashi} S. \& { Me\'sza\'ros} P., 2003, ApJ, 595, 950

\bibitem[]{zhang02} Zhang, B. \& M\'esz\'aros, P. 2002, ApJ, 581, 1236

\bibitem[]{zhang04} Zhang, B. \& M\'esz\'aros, P. 2004, IJMPA, 19, 2385

\bibitem{} Zhang, B. \& Yan, H. 2011, ApJ, 726, 90

\bibitem[{{Zou} {et~al.}(2005)}] {zou05} {Zou}, Y., Wu, X. \& Dai, Z., 2005, MNRAS, 363, 93

\end{thebibliography}
\end{document}